\documentclass[twocolumn,letterpaper,nofootinbib,prd,amsmath,superscriptaddress]{revtex4}

\usepackage[dvips]{graphics,color}
\usepackage{epsfig}
\usepackage{hyperref}

\newcommand{\fder}[2]{\ensuremath{\frac{d #1}{d #2}}}

\newcommand{\fpder}[2]{\ensuremath{\frac{\partial #1}{\partial #2}}}

\newcommand{\coder}[2]{\ensuremath{\nabla_{#2} #1}}
\newcommand{\conder}[2]{\ensuremath{\nabla^{#2} #1}}

\newcommand{\td}[2]{\ensuremath{#1_{#2}}}

\newcommand{\tdd}[3]{\ensuremath{#1_{#2 #3}}}
\newcommand{\tuu}[3]{\ensuremath{#1^{#2 #3}}}
\newcommand{\tud}[3]{\ensuremath{#1^{#2}_{\ #3}}}

\newcommand{\bssn}[1]{\ensuremath{\tilde{#1}}}
\newcommand{\hcoder}[2]{\ensuremath{\bssn{\nabla}_{#2} #1}}
\newcommand{\hconder}[2]{\ensuremath{\bssn{\nabla}^{#2} #1}}
\newcommand{\hChrist}[3]{\ensuremath{\bssn{\Gamma}^{#1}_{#2 #3}}}
\newcommand{\hChristd}[3]{\ensuremath{\bssn{\Gamma}_{#1 #2 #3}}}
\newcommand{\cChrist}[1]{\ensuremath{\bssn{\Gamma}^{#1}}}

\newcommand \RR {\ensuremath{\mathbf{R}}}
\newcommand \M {\ensuremath{\cal{M}}}
\newcommand \del {\ensuremath{\partial}}

\begin{document}

\title{Evolving a puncture black hole with fixed mesh refinement}

\date\today

\author{Breno Imbiriba}
\affiliation{Laboratory for High Energy Astrophysics,
NASA Goddard Space Flight Center, Greenbelt, MD 20771 USA}
\affiliation{Department of Physics, University of Maryland,
College Park, MD 20740 USA}
\author{John Baker}
\affiliation{Laboratory for High Energy Astrophysics,
NASA Goddard Space Flight Center, Greenbelt, MD 20771 USA}
\author{Dae-Il Choi}
\affiliation{Laboratory for High Energy Astrophysics,
NASA Goddard Space Flight Center, Greenbelt, MD 20771 USA}
\affiliation{Universities Space Research Association,
7501 Forbes Boulevard \#206, Seabrook, MD 20706 USA}
\author{Joan Centrella}
\affiliation{Laboratory for High Energy Astrophysics,
NASA Goddard Space Flight Center, Greenbelt, MD 20771 USA}
\author{David R. Fiske}
\affiliation{Department of Physics, University of Maryland,
College Park, MD 20740 USA}
\affiliation{Laboratory for High Energy Astrophysics,
NASA Goddard Space Flight Center, Greenbelt, MD 20771 USA}
\author{J.~David Brown}
\affiliation{Department of Physics, North Carolina State University,
Raleigh, NC 27695 USA}
\author{James R. \surname{van Meter}}
\affiliation{Laboratory for High Energy Astrophysics,
NASA Goddard Space Flight Center, Greenbelt, MD 20771 USA}
\author{Kevin Olson}
\affiliation{Goddard Earth Sciences and Technology Center,
University of Maryland Baltimore County, Baltimore, MD 21250 USA}
\affiliation{Earth and Space Data Computing Division,
NASA Goddard Space Flight Center, Greenbelt, MD 20771 USA}

\begin{abstract}
We present an algorithm for treating mesh refinement interfaces in numerical
relativity.  We discuss the behavior of the solution near such interfaces
located
in the strong field regions of dynamical black hole spacetimes,
with particular attention to the  convergence properties of the simulations.
In our applications of this technique
to the evolution of puncture initial data with vanishing shift, we 
 demonstrate that it is possible
to simultaneously maintain second order convergence near the puncture and
extend the outer boundary beyond $100M$, thereby approaching the asymptotically
flat region in which boundary condition problems are less difficult and
wave extraction is meaningful.
\end{abstract}

\maketitle

\section{Introduction}
\label{sec:intro}

Numerical relativity, which comprises the solution of Einstein's equations
on a computer, 
is an essential tool for understanding the behavior of strongly
nonlinear dynamical gravitational fields.  Current grid-based formulations of
numerical relativity feature $\sim 17$ or more coupled nonlinear partial
differential equations that are solved using finite differences in 3 spatial
dimensions (3-D) plus time.  The physical systems described by these equations
generally have a wide range of length and time scales, and realistic 
simulations are expected to require the use of some type of adaptive gridding 
in the spacetime domain.

A primary example of the type of physical system to be studied using numerical 
relativity is the final merger of two inspiraling black holes, 
which is expected to be a 
strong source of gravitational radiation for ground-based detectors such 
as LIGO and VIRGO, as well as the space-based LISA \cite{Schutz03}.
  The individual
black hole masses
$M_1$ and $M_2$ set the scales for the binary in 
the source interaction region, and we can expect both spatial and 
temporal changes on these scales as the system evolves.
The binary must 
be evolved for a time $t \sim 1000M$, $M \sim M_1 + M_2$,
starting from an orbital separation 
$\sim 10 M$
to simulate its final few orbits followed by  
the plunge and ringdown.  This orbital region is
surrounded by the wave zone with features of scale $\sim 100M$, 
where the outgoing
signals take on a wave-like character and can be measured.  Accomplishing
realistic simulations of binary black hole mergers on even the most
powerful computers clearly requires the use of variable mesh sizes over
the spatial grid.

Adaptive mesh refinement (AMR) was first applied in
numerical relativity to study critical phenomena in
scalar field collapse in 1-D \cite{Choptuik93};
several other related studies have also 
used AMR, most recently in 2-D \cite{Choptuik03}.
  AMR has also been used 
in 2-D to study the evolution of inhomogeneous
cosmologies \cite{Hern:1999dq,Belanger01}.
In the area of black hole evolution, AMR was first applied 
to a simulation of a Schwarzschild black hole
\cite{Bruegmann96}.
Fixed mesh refinement (FMR) was used to evolve a
 short part of a (nonequal mass) binary
black hole merger \cite{Bruegmann97},
an excised Schwarzschild black hole in an evolving
gauge \cite{CarpetFMR}, and orbiting, equal mass black holes in a 
co-rotating gauge \cite{Bruegmann03}.
AMR has also been used to set binary black hole initial data
\cite{Diener99,BrownLowe03}. 
The propagation of gravitational waves through spacetime has 
been carried out using AMR, first using a single 3-D model equation 
describing perturbations of a Schwarzschild black hole 
\cite{Papadopoulos98a}
and later in the 3-D Einstein equations \cite{New:2000dz}.  
Gravitational waves have also been propagated across fixed mesh
refinement boundaries, with a focus on the interpolation conditions
needed at the mesh boundaries to inhibit spurious reflected waves
\cite{Choi:2003ba}.  

Realistic simulations of the final merger of binary black holes are
likely to require a hierarchy of grids, using both FMR and AMR.  
The source region would have the finest grids, and would be surrounded
by successively coarser
grids, encompassing the orbital region and extending into the
wave zone out to distances $> 100 M$.  
Evolving dynamical gravitational fields using such a mesh
refinement hierarchy poses a number of technical challenges.  For 
example, the gravitational waves produced by the sources will 
originate as signals in the near zone and need to
cross fixed mesh refinement boundaries to reach the wave zone.  
  In addition, Coulombic-like 
 signals
 that may vary with time but are not wavelike in character,
such as are produced by the gravitational potential around black
holes, can stretch across mesh boundaries. 
Inappropriate interpolation conditions at refinement boundaries can
lead to spurious reflection of signals at these interfaces; 
\emph{cf.}~\cite{Choi:2003ba}.
   Additional complications can arise when the grid
refinement is adaptive.

In this paper, we use the evolution of a single Schwarzschild black hole
with FMR as
a numerical laboratory. We represent the black hole as a puncture 
without excision and use gauges with zero shift
in which the solution undergoes significant 
evolution.  This tests the ability of our code to handle dynamically
changing spacetimes in the vicinity of mesh refinement
boundaries. Using  a hierarchy
of fixed mesh refinements, we are able to
resolve the strong field region near the puncture (and demonstrate
the convergence of the solution in this region) while locating the
outer boundary at $> 100M$.  
In Sec.~\ref{sec:methodology} we describe our methodology,
including the numerical implementation.  The treatment of mesh
refinement boundaries is discussed in Sec.~\ref{sec:interfaces}.
Black hole evolutions with FMR are presented in 
Sec.~\ref{sec:BHevol}; examples are given of evolutions using
geodesic slicing,
and 1 + log slicing with zero shift.
  We conclude with a summary in
Sec.~\ref{sec:summary}.

\section{Methodology}
\label{sec:methodology}

\subsection{Basic Equations}
\label{sec:eqns}
We use the BSSN form of the ADM equations \cite{Shibata95,Baumgarte99}.  
These equations evolve the quantities
\begin{subequations}\label{eqn:BSSNdefs}
\begin{eqnarray}
\phi & = & \frac{1}{12} \log \gamma \label{eqn:BSSNphi} \\
K & = & \tuu{\gamma}{a}{b} \tdd{K}{a}{b} \label{eqn:BSSNK} \\
\tdd{\bssn{\gamma}}{i}{j} & = & e^{-4 \phi} \tdd{\gamma}{i}{j} 
	\label{eqn:BSSNg} \\
\tdd{\bssn{A}}{i}{j} & = & e^{-4 \phi} \left( 
	\tdd{K}{i}{j} - \frac{1}{3} \tdd{\gamma}{i}{j} K
\right) \label{eqn:BSSNA} \\
\cChrist{i} & = & \tuu{\bssn{\gamma}}{a}{b} \hChrist{i}{a}{b} 
	\label{eqn:BSSNGamma}
\end{eqnarray}
\end{subequations}
written here in terms of the physical, spatial 3-metric 
\tdd{\gamma}{i}{j} and extrinsic curvature \tdd{K}{i}{j} \cite{York79},
where all indices range from 1 to 3.  In Eq.~(\ref{eqn:BSSNGamma}),
$\tilde{\Gamma}^{i}_{ab}$ is the Christoffel symbol associated with the 
conformal metric $\tilde{\gamma}_{ij}$.
These quantities evolve according to
\begin{subequations}\label{eqn:BSSN}
\begin{eqnarray}
\fder{\phi}{t} & = & -\frac{1}{6} \alpha K \label{eqn:phievol} \\
\fder{K}{t} & = & -\conder{\coder{\alpha}{a}}{a}
	+ \alpha \left( \tdd{\bssn{A}}{a}{b} \tuu{\bssn{A}}{a}{b}
		+ \frac{1}{3} K^{2}
	\right) \label{eqn:trKevol} \\
\fder{\tdd{\bssn{\gamma}}{i}{j}}{t} & = & -2 \alpha \tdd{\bssn{A}}{i}{j} 
	\label{eqn:gevol} \\
\fder{\tdd{\bssn{A}}{i}{j}}{t} & = &
	e^{-4\phi} \left( -\coder{\coder{\alpha}{j}}{i}
	+ \alpha \tdd{R}{i}{j} \right)^{\mathrm{TF}} \nonumber \\
& & \mbox{}
	+ \alpha \left( 
		K\tdd{\bssn{A}}{i}{j} 
		- 2\tdd{\bssn{A}}{i}{a}\tud{\bssn{A}}{a}{j} 
	\right) \label{eqn:Aevol} \\
\fpder{\cChrist{i}}{t} & = &
	2\alpha \left( \hChrist{i}{a}{b} \tuu{\bssn{A}}{a}{b}
		- \frac{2}{3} \tuu{\bssn{\gamma}}{i}{a} \td{K}{,a}
		+ 6 \tuu{\bssn{A}}{i}{a} \td{\phi}{,a}
	\right)	 \nonumber \\
& & \mbox{}
	+ \bssn{\gamma}^{kl} 
	\left( -\hChrist{j}{k}{l} \beta^{i}_{\ ,j}
	+ \frac{2}{3} \hChrist{i}{k}{l} \beta^{j}_{\ ,j} \right)
	+ \beta^{k}\cChrist{i}_{\ ,k} \nonumber \\
& & \mbox{}
	+ \bssn{\gamma}^{jk} \beta^{i}_{\ ,jk} 
	+ \frac{1}{3} \bssn{\gamma}^{ij} \beta^{k}_{\ ,kj}
	- 2 \tuu{\bssn{A}}{i}{a} \td{\alpha}{,a} \label{eqn:Gammaevol}
\end{eqnarray}
\end{subequations}
where here and henceforth the indices of conformal quantities are
raised with the conformal metric.
The lapse $\alpha$ and shift $\beta^{i}$ specify the gauge, the full
derivative notation 
\(d/dt = \partial/\partial t - \mathcal{L}_{\beta}\)
is a partial with respect to time minus a Lie derivative, and the notation
``TF'' indicates the trace-free part of the expression in parentheses.
These quantities are analytically subject to the conditions
\begin{subequations}\label{eqn:physConstraints}
\begin{eqnarray}
H & = & R - K_{ab}K^{ab} + K^{2} = 0 \label{eqn:ham} \\
P^{i} & = & (\gamma^{im}\gamma^{jn} - \gamma^{ij}\gamma^{mn}) 
\nabla_{j}K_{mn} = 0, \label{eqn:mom}
\end{eqnarray}
\end{subequations}
known respectively as the Hamiltonian and momentum constraints.
When evaluating Eqs.~(\ref{eqn:physConstraints})
we recompute the physical quantities
from the evolved quantities using Eqs.~(\ref{eqn:BSSNdefs}).  

Note that the covariant derivatives of the lapse in Eqs.~(\ref{eqn:trKevol})
and~(\ref{eqn:Aevol}) are with respect to the \emph{physical} metric, 
and are used here for compactness.  In the code, this is computed 
according to
\begin{eqnarray}
\coder{\coder{\alpha}{n}}{m} & = & \partial_{m} \partial_{n} \alpha
- 4 \partial_{(m} \phi \partial_{n)} \alpha \nonumber \\
& & \mbox{} - \hChrist{k}{m}{n} \partial_{k} \alpha 
+ 2 \tdd{\bssn{\gamma}}{m}{n} \tuu{\bssn{\gamma}}{k}{l} \partial_{k} 
\phi \partial_{l} \alpha
\label{eqn:coderalpha}
\end{eqnarray}
using
only the conformal BSSN quantities, and the index of the covariant 
derivative is raised on the right hand side of
Eq.~(\ref{eqn:trKevol}) with the physical metric.  The Ricci tensor in
Eq.~(\ref{eqn:Aevol}) is also with respect to the physical metric.  We 
compute it according to the decomposition
\begin{equation}
R_{ij} = \tilde{R}_{ij} + R^{\phi}_{ij} 
\end{equation}
with
\begin{eqnarray}
\bssn{R}_{ij} & = & -\frac{1}{2} \bssn{\gamma}^{lm} \bssn{\gamma}_{ij,lm}
	+ \bssn{\gamma}_{k(i} \bssn{\Gamma}^{k}_{\ ,j)} 
	+ \bssn{\gamma}^{lm} \hChrist{k}{l}{m} \hChristd{(i}{j)}{k}
	\nonumber \\
 & & \mbox{} + \bssn{\gamma}^{lm} \left( 
	2\hChrist{k}{l}{(i}\hChristd{j)}{k}{m} 
	+ \hChrist{k}{i}{m}\hChristd{k}{l}{j} \right) \label{eqn:confRicci}
\end{eqnarray}
and
\begin{eqnarray}
R^{\phi}_{ij} & = & -2 \hcoder{\hcoder{\phi}{j}}{i} 
     - 2 \bssn{\gamma}_{ij} \hconder{\hcoder{\phi}{k}}{k} \nonumber \\
 & & \mbox{} + 4 \hcoder{\phi}{i} \hcoder{\phi}{j} 
	- 4 \bssn{\gamma}_{ij} \hconder{\phi}{k} \hcoder{\phi}{k}
	\label{eqn:phiRicci}
\end{eqnarray}
giving the conformal and remaining pieces of the physical Ricci tensor.  The
notation $\tilde{\nabla}_{i}$ denotes the covariant derivative associated with
the conformal metric.

There are many rules of thumb in the community regarding how to incorporate
the constraints into the evolution equations, and, in particular, when to
use the independently evolved $\bssn{\Gamma}^{i}$ as opposed to recomputing
the equivalent quantity from the evolved metric.  We have made our choices
manifest in the writing of the equations here; we largely follow the
rules set out in \cite{Alcubierre02a}.

\subsection{Numerical Implementation}
\label{sec:numerics}
For the spatial discretization of Eqs.~(\ref{eqn:BSSN}),
we take the data to be defined at the centers of the spatial grid cells
and use standard $O(\Delta x)^2$ centered spatial differences \cite{Press86}.
To advance this system in time, we use 
the iterated Crank--Nicholson (ICN) method with 2 iterations \cite{Teukolsky00}, 
which gives $O(\Delta t)^2$ accuracy.  
We employ interpolated Sommerfeld outgoing
wave conditions at the outer boundary \cite{Shibata95} on all variables, 
except for the $\tilde{\Gamma}^{i}$ which are kept
 fixed at the outer boundary.
Overall, the code is second-order
convergent; specific examples of this are given in 
Sec.~\ref{sec:BHevol} below.

We explicitly enforce the algebraic constraints
that $\bssn{A}_{ij}$ is trace-free and that ${\tilde \gamma} = 1$ after
each ICN iteration.  We
enforce the trace-free condition by replacing the evolved variable with 
\begin{displaymath}
\bssn{A}_{ij} \rightarrow \bssn{A}_{ij} 
	- \frac{1}{3} \bssn{\gamma}^{mn} \bssn{A}_{mn} \bssn{\gamma}_{ij}
\end{displaymath}
and we enforce the unit determinant condition by replacing
the evolved metric with
\begin{displaymath}
\bssn{\gamma}_{ij} \rightarrow \bssn{\gamma}^{-1/3} \bssn{\gamma}_{ij}\ .
\end{displaymath}
Both of these constraints are
enforced in all of the runs presented below.
Since the $\cChrist{i}$ are evolved as independent quantities,
Eq.~(\ref{eqn:BSSNGamma}) acts as a further constraint on this system
of equations.  We monitor the behavior of this so-called 
$\cChrist{i}$ constraint along with the Hamiltonian and momentum
constraints, Eqs.~(\ref{eqn:ham}) and~(\ref{eqn:mom})

We use the Paramesh package \cite{paramesh,parameshMan}
 to implement both  mesh refinement and parallelization in
our code.  Paramesh works on logically Cartesian, or structured, 
grids and carries out
the mesh refinement on grid blocks.  When refinement is needed, the
grid blocks needing refinement are bisected in each coordinate direction, 
similar to the technique of Ref. \cite{deZP}.

All grid blocks have the same logical structure, with $n_{x}$ zones in
the $x$-direction, and similarly for $n_{y}$ and $n_{z}$. Thus, 
refinement of a 3-D block produces
 eight child blocks, each having $n_{x} n_y n_z$ zones but with zone
sizes in each direction a factor of two smaller than in the parent block. 
Refinement can continue as needed 
on the child blocks, with the restriction that the 
grid spacing can change only by a factor of two, or one refinement level, 
at any location in the spatial domain.  
Each grid block is surrounded by a number of guard cell layers that are used to
calculate finite difference spatial derivatives near the block's boundary. 
These guard 
cells are filled using data from the interior cells of the given block and the 
adjacent block; see Sec.~\ref{sec:interfaces}.

Paramesh can be used in applications requiring FMR, AMR, or a combination of these. 
The package takes care of creating the  grid blocks, as well as building
 and
maintaining 
 the data structures needed to track the spatial relationships between blocks.
Paramesh handles all inter-block communications and keeps track of physical boundaries
on which particular conditions are set, guaranteeing that the child blocks inherit
this information from the parent blocks.  In a parallel environment, Paramesh distributes
the blocks among the available processors to achieve load balance, 
minimize inter-processor communications, and maximize block locality.

This scheme provides excellent computational scalability. 
Equipped with Paramesh,
the scalability of our code has been tested for up to 256 processors 
and has demonstrated
a consistently good scaling factor
for both unigrid (uniform grid) and FMR runs. 
For unigrid runs, we started with a uniform Cartesian grid
of a certain number of grid cells, a fixed number of timesteps,
and a certain number of PEs (Processing Elements), and then
increased the number of PEs to run a larger job 
while the number of grid cells per PE remained constant.
In this situation, we expect that the total time taken 
to run the code, including the CPU time used by all of the 
PEs, should scale linearly with the size of
the problem under perfect conditions.
In reality, communication overhead makes the scalability less than
perfect.
We define the scaling factor to be the time expected with
perfect scaling divided by the actual time taken.
Using FMR, we ran the same simulations as in the unigrid case except
that
a quarter of the computational domain was covered by a mesh with
twice the resolution.
Despite the more complicated communication patterns,
scalability in the FMR runs is comparable to that in the unigrid runs.
The scaling factor of our code is 0.92 for 
unigrid runs and 0.90 for FMR runs.

For the work described in this paper, we are using FMR.  
For simplicity, we use the same timestep, chosen for stability on
the finest grid and with a Courant factor of 0.25,
 over the entire computational domain;
\emph{cf.}~\cite{DursiZingale}.
  At the mesh refinement
boundaries, we use a single layer of guard cells. 
Special attention is paid to
the restriction (transfer of data from fine to coarse grids) and 
prolongation (coarse
to fine) operations used to set the data in these guard cells, as 
discussed in the next
section. 

\section{Treatment of Refinement Boundaries}
\label{sec:interfaces}
Careful treatment of guard cells at mesh refinement boundaries is
needed to produce accurate and robust numerical simulations.
The current 
version of our code uses a third 
order\footnote{In our terminology 
the  ``order of accuracy'' refers to the order of errors in the grid spacing. 
Thus, third order accuracy for guard cell filling means that the guard cell 
values have errors of order $\Delta x^3$, where $\Delta x$ is the (fine) grid 
spacing.  (Note that third order accurate guard cell filling was termed 
``quadratic'' guard cell filling in Ref.~\cite{Choi:2003ba}.) Second order 
accuracy for the evolution code means that, after a finite evolution time, 
the field variables have errors of order $\Delta x^2$.}
guard cell filling scheme that is now 
included with the standard Paramesh package. This guard cell filling proceeds 
in three steps. 

\begin{figure*}
\centerline{\includegraphics{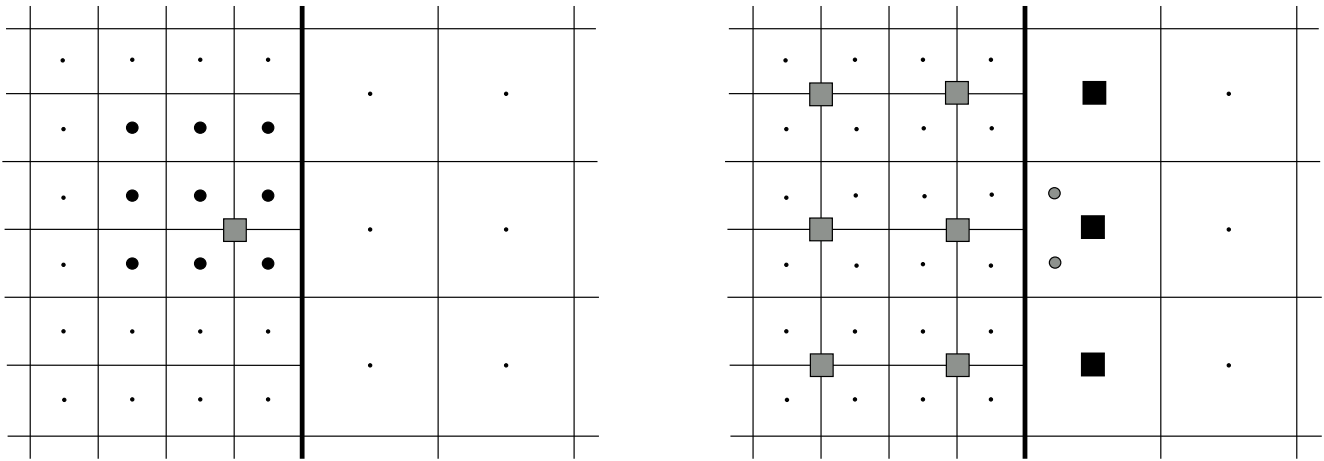}}
\caption{Guard cell filling in two spatial dimensions. In these pictures, 
the thick vertical line represents a refinement boundary separating 
fine and coarse grid regions.   
The picture on the left shows the first step, in which one of the parent grid 
cells (grey square) is filled using quadratic interpolation 
across nine interior fine grid cells (black circles). 
The other parent grid cells are filled using corresponding stencils of 
nine interior fine grid cells. (The asymmetry in the left panel 
is drawn with the assumption that the fine block's center is 
toward the top-left of the panel.)
The picture on the right shows the second step in which 
two fine grid guard cells (grey circles) are filled using 
quadratic interpolation across nine parent grid values (squares). 
These parent grid values include one layer of guard cells (black squares) obtained from 
the coarse grid region to the right of the interface, and two layers of interior cells (grey squares). 
The final step in guard cell filling (not shown in this figure) is to use ``derivative matching'' to 
fill the guard cells for the coarse grid.}
\label{GCFfigure}
\end{figure*}
The first step is a restriction operation in which interior fine grid cells 
are used to fill the interior grid cells of the underlying ``parent'' grid.  
The parent grid is a grid that covers the same domain as the 
fine grid but has twice the grid spacing. 
The restriction operation is depicted for the case of two spatial dimensions 
in the left panel of 
Fig.~\ref{GCFfigure}.

The restriction proceeds as a succession of one--dimensional quadratic 
interpolations, and is accurate to third order in the grid spacing. 
Note that the 3-cell-wide fine grid stencil used for this step 
(nine black circles in the figure) cannot be centered on the parent cell
(grey square).  In each dimension the stencil includes two fine 
grid cells on one side of the parent cell and one fine grid cell on the other. 
The stencil is always positioned so that its center is shifted toward 
the center of the block (assumed in the figure to be toward the upper left).  
This ensures that only interior fine grid points, and no fine grid guard cells,
are used in this first step.

For the second step, the fine grid guard cells are filled by prolongation 
from the parent grid. Before the prolongation, the parent grid gets 
its own guard cells (black squares in the right panel of Fig.~\ref{GCFfigure}) 
from the neighboring grids of the same refinement level, 
in this case from the coarse grid. 
The stencil used in the prolongation operation is shown in the right panel 
of Fig.~\ref{GCFfigure}.
The prolongation operation proceeds as a succession of one--dimensional 
quadratic interpolations, and is third order accurate. 
In this case, the parent grid stencil includes a layer of guard cells 
(black squares), 
as well as its own interior grid points (grey squares).
At the end of this second step the fine grid guard cells are filled 
to third order accuracy. 

The third step in guard cell filling is ``derivative matching'' 
at the interface.\footnote{In Ref.~\cite{Choi:2003ba} this process is
referred to as ``flux matching.''}  
With derivative matching the coarse grid guard cell 
values are computed so that the first derivatives at the interface, 
as computed on the coarse grid, match the first derivatives at the interface 
as computed on the fine grid. The first derivative on the coarse grid is 
obtained from standard second order differencing using a guard cell 
and its neighbor across the interface. The first derivatives on the fine grid
are computed using guard cells and their neighbors across the interface, 
appropriately averaged to align with the coarse grid cell centers. 
This third step fills the coarse grid guard cells to third order 
accuracy. 

An alternative to derivative matching, which we do not use, is to 
fill the coarse grid guard cells from the first layer of interior cells 
of the parent grid. However, we find that such a scheme leads to unacceptably 
large reflection and transmission errors for waves passing 
through the interface.
These errors are suppressed by derivative matching.

Why should third order guard cell filling be adequate to maintain 
overall second order accuracy? 
This is a nontrivial question, and there are certain subtleties 
that arise in our black hole evolutions. In Appendix~\ref{sec:errors}, 
we present a detailed 
error analysis for our guard cell filling algorithm
based on simplified model equations for a scalar field 
in 1-D. This toy model shares many of the features of the full
BSSN system and provides a useful guide to understanding the behavior
of our black hole evolutions.  We demonstrate that, with this
algorithm, second spatial 
derivatives of the BSSN variables defined by Eqs.~(\ref{eqn:BSSNdefs})
acquire first 
order errors at grid points adjacent to mesh refinement boundaries. 
These first order errors show up as spikes in a convergence plot for 
quantities that depend on second spatial derivatives, 
such as the Hamiltonian constraint. The key result of this analysis, however,
is a demonstration that the first order errors 
in second derivatives do not spoil the overall second order 
convergence of the evolved variables in Eqs.~(\ref{eqn:BSSNdefs}), 
in spite of the fact that second spatial derivatives 
appear on the right-hand sides of the evolution equations (\ref{eqn:BSSN});
 {\emph{cf.} \cite{Chesshire90,Henshaw03}.  

\section{Black Hole Evolutions}
\label{sec:BHevol}
Black hole spacetimes are a particularly challenging subject for numerical 
study.  Astrophysical applications will require that our FMR
implementation perform robustly under the adverse conditions which
arise in black hole simulations, 
such as strong time-dependent potentials
and propagating signals that become gravitational waves.  
In this section we demonstrate that
our techniques perform convergently and accurately in the presence of
strong field dynamics and singular ``punctures'' associated with
black hole evolutions, and that these methods can be stable on the
timescales required for interesting simulations. 

The puncture approach to black hole spacetimes generalizes the
Brill-Lindquist \cite{Brill63} prescription of initial data for black holes
at rest.  In this approach, the spacetime is sliced in such a way
as to avoid intersecting the black hole singularity, and the spatial
slices are topologically isomorphic to $\RR^3$ minus one point, a
puncture, for each hole.  The punctures represent an inner asymptotic
region of the slice which can be conformally transformed to data which
are regular on $\RR^3$.  In this way a resting black hole of mass $M$
located at $r=0$ is expressed in isotropic spatial coordinates by
$\gamma_{ij}=\Psi^4_{BL}\delta_{ij}$, with 
conformal factor $\Psi_{BL}=1+{M}/{2r}$
and $ K_{ij}=0$. A direct generalization of this expression for the
conformal factor can be used to represent multiple black hole
punctures, and data for spinning and moving black holes can be
constructed according to the Bowen-York \cite{Bowen80} prescription.

A key characteristic which makes this representation appropriate for
spacetime simulations is that with suitable conditions on the
regularity of the lapse and shift, the evolution equations imply that
time derivatives of the data at the puncture are regular everywhere
despite the blow up in $\Psi_{BL}$ at the puncture.  Numerically, we
treat the punctures by a prescription similar to that given in
\cite{Alcubierre02a}.  In the BSSN formulation, this amounts to a
splitting of the conformal factor $\exp({4\phi})$ into a regular part, $\exp({4\phi_r})$, and
non-evolving singular part, $\exp({4\phi_s})$, given by 
$\Psi_{BL}$.  The numerical grid is staggered to make sure the
puncture does not fall directly on a grid point, and to avoid the
large finite differencing error, the derivatives of $\phi_s$ are
specified analytically.
 
We study two test problems, each representing a Schwarzschild black
hole in a different coordinate system.  
Both problems test the performance of our FMR interfaces under the
condition that strong-field spacetime features pass through the
interfaces.
The first case, described in
Sec.~\ref{sec:geodesic}, is a black hole in geodesic
coordinates, in which the data evolve as the slice quickly advances
into the singularity.  In Appendix~\ref{sec:analytic},
we present an analytic solution for the
development of this spacetime with which we can compare for a direct test
of the simulation.  Our next test, in Sec.~\ref{sec:1+log_results},
uses a variant of the ``1+log'' slicing
condition to define the lapse, $\alpha$, with vanishing shift,
$\beta^i=0$. This gauge choice allows the slice to avoid running into the
singularity, but causes the black hole to appear to grow in coordinate
space so that the horizon passes though our FMR interfaces.
\begin{figure}
\includegraphics[scale=.7,angle=-90]{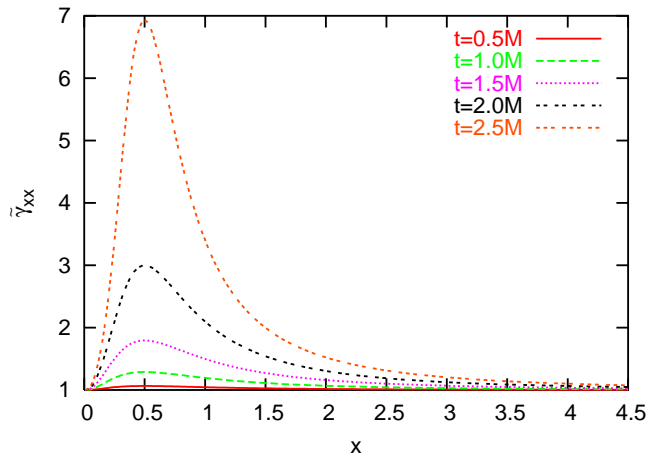}
\caption{Evolution of the conformal metric component
$\tilde \gamma_{xx}$, for a geodesically sliced puncture, shown at 
$t=0.5$, 1.0, 1.5, 2.0, and $2.5 M$.}
\label{fig:g-evol}
\end{figure}

\subsection{Geodesic Slicing}
\label{sec:geodesic}

We begin with the numerical evolution of a single puncture black hole
using geodesic slicing.
\begin{figure*}
\hspace*{\fill}
\includegraphics[scale=.7,angle=-90]{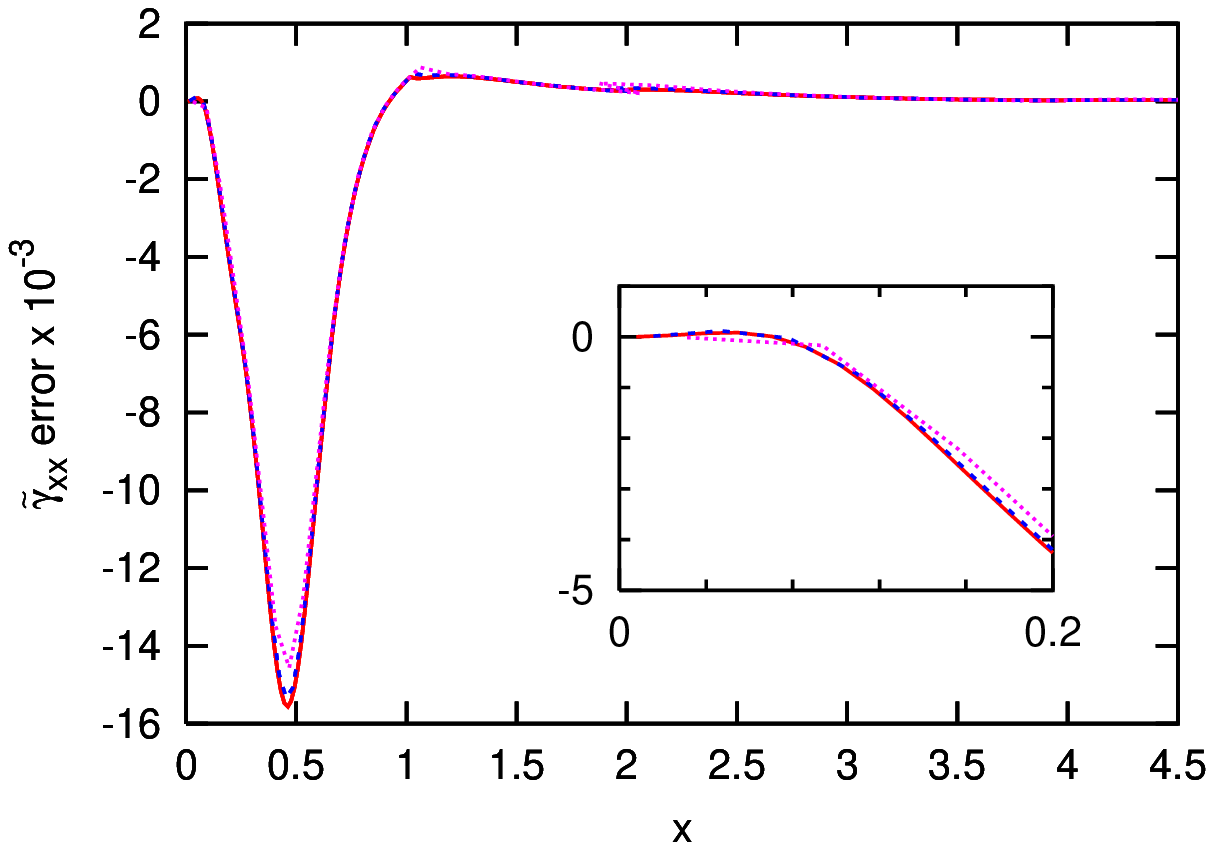}
\hspace*{\fill}
\includegraphics[scale=.7,angle=-90]{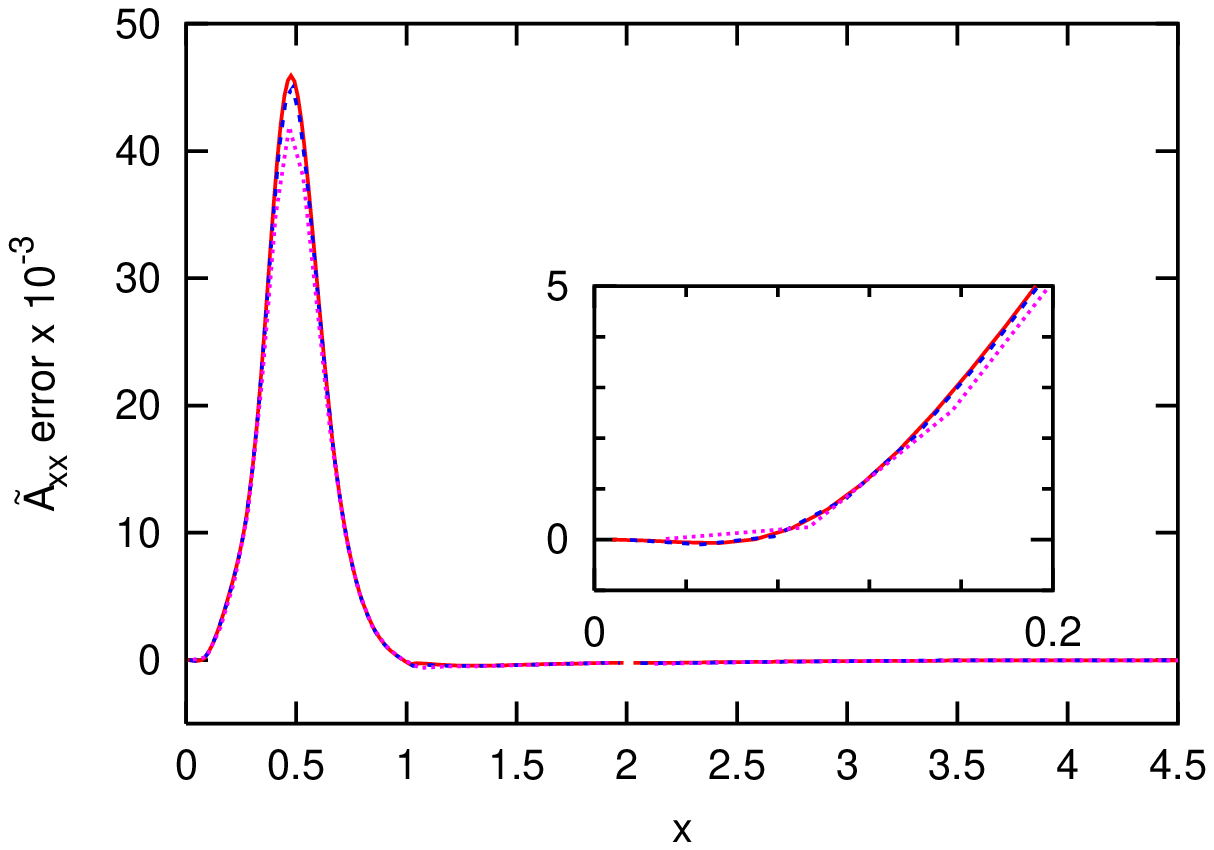}
\hspace*{\fill}
\\
\hspace*{\fill}
\includegraphics[scale=.7,angle=-90]{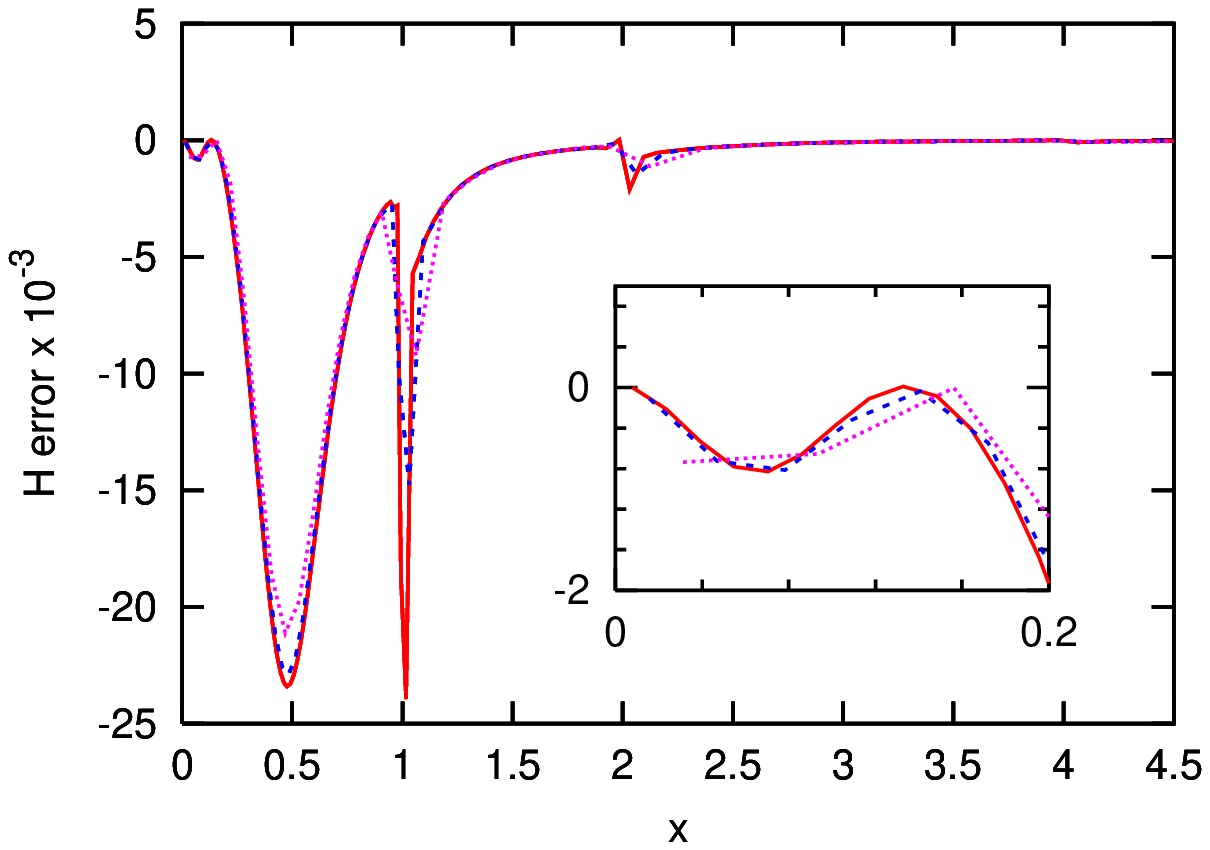}
\hspace*{\fill}
\includegraphics[scale=.7,angle=-90]{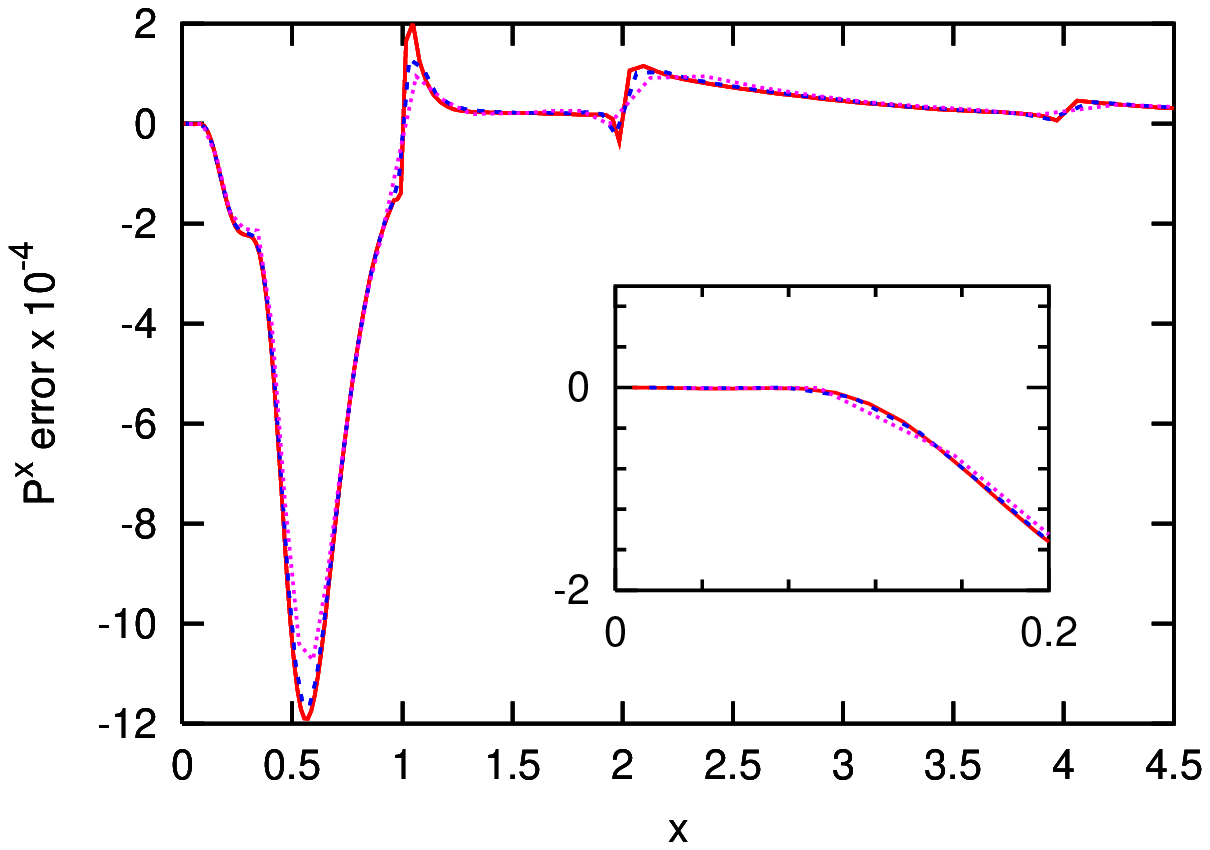}
\hspace*{\fill}
\caption{Convergence of the errors (numerical values minus analytic
values) in $\tilde \gamma_{xx}$, $\tilde A_{xx}$, the Hamiltonian constraint 
$H$, and the momentum constraint $P^{x}$, for a geodesically sliced puncture
along the $x$-axis,
all at the time $t=2.5 M$.  The solid line shows the errors for the highest
resolution run.  The errors for the medium resolution run (dashed line)
and the lowest resolution run (dotted line) have been divided by factors
of 4 and 16, respectively, to demonstrate second order convergence.
 Note that the full domain of the simulation extends
to $128M$.}
\label{fig:gKHP-conv}
\end{figure*}
As explained in Appendix~\ref{sec:analytic}, 
at $t=\pi M$
the slice $\Sigma_\pi$  on which our data resides 
will reach the physical singularity\footnote{In Appendix~\ref{sec:analytic},
we use $\tau$ as the time coordinate.  When using the results
derived there in the main body of this paper, we
relabel $\tau \rightarrow t$ to conform to the notation set out
in Sec.~\ref{sec:methodology}}; because we are not 
performing any excision, this sets the maximum duration of our evolution. 
Nevertheless, $t\approx 3M$ is long enough to test the relevant features of 
FMR evolution and provide us with a simple analytical solution.

In these simulations we locate the puncture black hole at the
origin. The spherical symmetry of the problem allows us to restrict
the simulation to an octant domain with symmetry boundary conditions,
thereby saving memory and computational time. The outer
boundaries are planes of constant $x$, $y$, or $z$ at $128M$ each.  As noted
before, one of the important uses of FMR is to enable the outer
boundary to be very far from the origin, at 
$\gtrsim 100M$. In this case we can
apply our exact solution as the outer boundary condition, though 
any numerical effects produced at this
distant boundary are completely irrelevant for the most interesting
strong-field region.  

 In this test we are mainly interested in how the FMR boundaries
behave near the puncture and under strong gravitational fields.  Even
with the outer boundary far away we can, by applying multiple nested
refinement regions, highly resolve the region near the puncture, as is
required to demonstrate numerical convergence.  To achieve the desired
resolution near the puncture, we use 8 cubical refinement levels,
locating the refinement boundaries at the planes 
$64M$, $32M$, $16M$, $8M$, $4M$, $2M$ and
$1M$ in the $x$-, $y$-, and $z$-directions. 

To test convergence we will examine the results of three runs with
identical FMR grid structures, but different resolutions. The lowest
resolution run has gridpoints $\Delta x_f = \Delta y_f = \Delta z_f = M/16 $ apart in the finest refinement
region near the puncture.  The medium resolution run has double the
resolution of the first run in each refinement region. The 
highest resolution run has twice the resolution of the medium
resolution run, for a maximum resolution of
$\Delta x_f = M/64$ near the puncture. 
The memory demand and computational load per timestep 
for the low, medium, and high resolution runs
are similar to unigrid runs of $32^3$, $64^3$ and $128^3$
gridpoints.

Since the data in our simulations are defined at the centers of
the spatial grid cells (see Fig.~\ref{GCFfigure}), we must interpolate
when extracting data on cuts through the simulation volume.  We use
cubic interpolation, which is accurate to order $\Delta x^4$, to 
insure that the interpolation errors are smaller than the largest
differencing errors of order $\Delta x^2$ expected in the simulations;
\emph{cf.}\ the discussion on
post-processing in \cite{Choptuik92b}.
When interpolating at a location near a refinement boundary, we adjust
the stencil so that the interpolation involves only data points
at the same level of refinement while still maintaining order
$\Delta x^4$ errors. 

In Fig.~\ref{fig:g-evol} we plot the conformal
metric component $\tilde \gamma_{xx}$ for the highest resolution run
along the $x$-axis at times 
$t=0.5$, 1.0, 1.5, 2.0,
and $2.5M$.  Note that in these coordinates the event horizon is at
$r = 0.5M$ at $t=0$ and moving outward 
toward larger values of the radial coordinate.  By $t=\pi M$,
the \emph{singularity} is at coordinate position $r = 0.5M$, so the mesh
refinement interface at $x=1M$ is truly in the strong field regime.
Because the slice will hit the singularity at coordinate position
$x =0.5M$, the metric grows sharply there as the simulation time advances.
 
In the present context
though, we are not so much interested in the field values of this
well-studied spacetime, as in the simulation errors,
that is, the differences between the analytical solution presented
in Appendix~\ref{sec:analytic} and the
numerical results.  These differences allow us to directly measure the 
errors in our numerical simulation.  At late times, these errors are, not 
surprisingly, dominated by
finite differencing errors in the vicinity of the developing singularity.

The plots in Fig.~\ref{fig:gKHP-conv}
compare these errors along the $x$-axis  at $t=2.5M$
for the three different resolution runs described above, demonstrating the 
convergence of $\tilde \gamma_{xx}$, $\tilde A_{xx}$, the Hamiltonian
constraint, and the $x$-component of the momentum constraint.
In each panel, the solid line shows the errors for the high resolution run.
The errors for the medium (dashed line) and low (dotted line)
resolution runs have been divided by 4 and 16, respectively.  That the curves
shown lie nearly atop one another is an indication of second order 
convergence, \emph{i.e.}\ that the lowest order error term depends
quadratically on the gridspacing, $\Delta x$. 
That the remaining difference between the
adjusted curves near $x = 0.5M$ seems also to decrease quadratically is an indication
that the next significant error term is of order $\Delta x^4$. 
We achieve convergence to the analytic solution everywhere, from very near 
$x = 0$, through the peak region which is approaching the singularity, 
and in the weak field region.  Animations of these results can be found
in the APS auxiliary archive EPAPS; see Fig.\ 3A and the associated
animation file in Ref.\ \cite{EPAPS-FMR}.

Our particular interest is in the region near the refinement
interfaces.  Fig.~\ref{fig:g-zoom} shows a
\begin{figure}
\includegraphics[scale=.7,angle=-90]{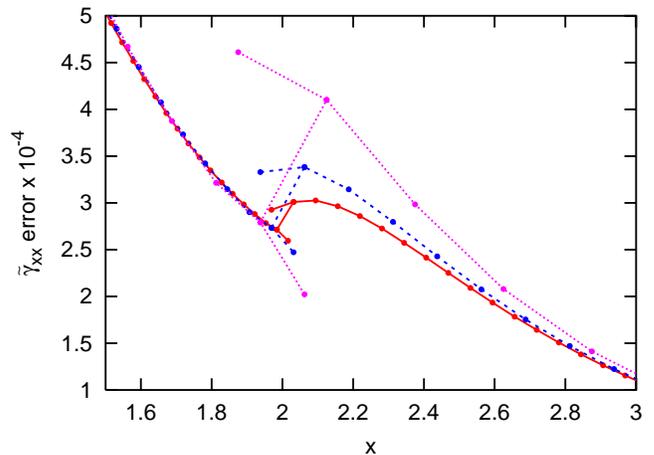}
\caption{Close-up of the convergence of the error (numerical value minus 
analytic value) in $\tilde \gamma_{xx}$, for a geodesically sliced puncture, 
at $t = 2.5M$, in the vicinity of the refinement boundary at $x = 2M$.
The errors for the high resolution run are shown by the solid line;
the errors for the medium (dashed line) and low (dotted line) resolution
runs have been divided by factors of 4 and 16, respectively. In this plot,
we also show the location of the data points, including guardcells,
using filled circles.}                               
\label{fig:g-zoom}
\end{figure}
close-up of $\tilde \gamma_{xx} $ near the refinement boundary at
$x=2M$.  In this figure we have included the values of the guardcells
used for defining finite-difference stencils near the interface. 
Again, the curves lie nearly atop one another, indicating second
order convergence. 

A similar close-up of
the Hamiltonian constraint $H$ requires a little more explanation. 
Eq.~(\ref{eqn:ham}) involves up to second derivatives of the
BSSN data, implemented by finite differencing. 
Fig.~\ref{fig:g-blowup} shows the
\begin{figure}
\includegraphics[scale=.7,angle=-90]{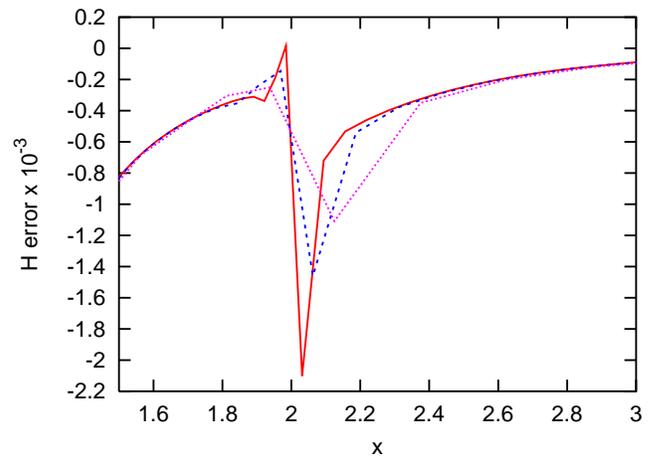}
\caption{A view of the Hamiltonian constraint in the neighborhood of the
FMR interface at $x=2M$. 
The errors for the high resolution run are shown by the solid line;
the errors for the medium (dashed line) and low (dotted line) resolution
runs have been divided by factors of 4 and 16, respectively.
 As discussed in the text, data points nearest to the 
interface converge at one order lower than in the rest of the domain.}
\label{fig:g-blowup}
\end{figure}
computed values of the Hamiltonian constraint in the vicinity of the
refinement boundary at $x=2M$.  While the result is second-order
convergent at any specific physical point in the neighborhood of the
boundary, the figure indicates that the sequence of values computed at the
nearest point approaching the interface as $\Delta x\rightarrow 0$
approaches zero at only first order.  This is as expected according to
the discussion in Appendix~\ref{sec:errors} (\emph{cf.}\
Fig.~\ref{psippfigure}) for a derived quantity involving second
derivatives.  We have specifically verified that, as with
$\tilde{\gamma}_{xx}$, all BSSN variables converge to second order at the
refinement boundaries.

We also examined the simulation data along cuts away from the
$x$-axis and have found them to be qualitatively similar to those
on the axis. In particular, plots and animations of the
errors along the line $y = z = 0.25M$ can be found in the
EPAPS supplement; see
Fig.\ 3B and the associated animation file in Ref.\ \cite{EPAPS-FMR}.
This particular 1-D cut
is instructive since it includes the strong field region yet has no
particular symmetric relation to the solution. The fact that the
errors along this line are qualitatively similar to those along the
$x$-axis gives us confidence that the
results we display in Fig.~\ref{fig:gKHP-conv} are not subject to
accidental cancellations due to octant symmetry boundary
conditions that might produce artificially small errors.

We have also examined the $L_1$ and $L_2$ norms of the errors in basic
variables and constraints to assess the overall properties of the simulation.
Representative results are shown in Fig.~\ref{fig:norms_geod}, where
the top panel displays the convergence behavior of the $L_2$ norm of
of the error in $\tilde\gamma_{xx}$ and the bottom panel the 
convergence of the $L_2$ norm of $H$.  The errors for the medium (dashed
line) and low (dotted line) resolution runs have been divided by 4 and 16, 
respectively.  These curves lie nearly atop the errors for the high
resolution run (solid line), indicating the second order convergence of
these error norms; see Appendix~\ref{sec:norms} and 
Eqs.~(\ref{eq:L1-scale-basic.vars}) and~(\ref{eq:L2-scale-basic.vars}).

\begin{figure}
\includegraphics[scale=.7,angle=-90]{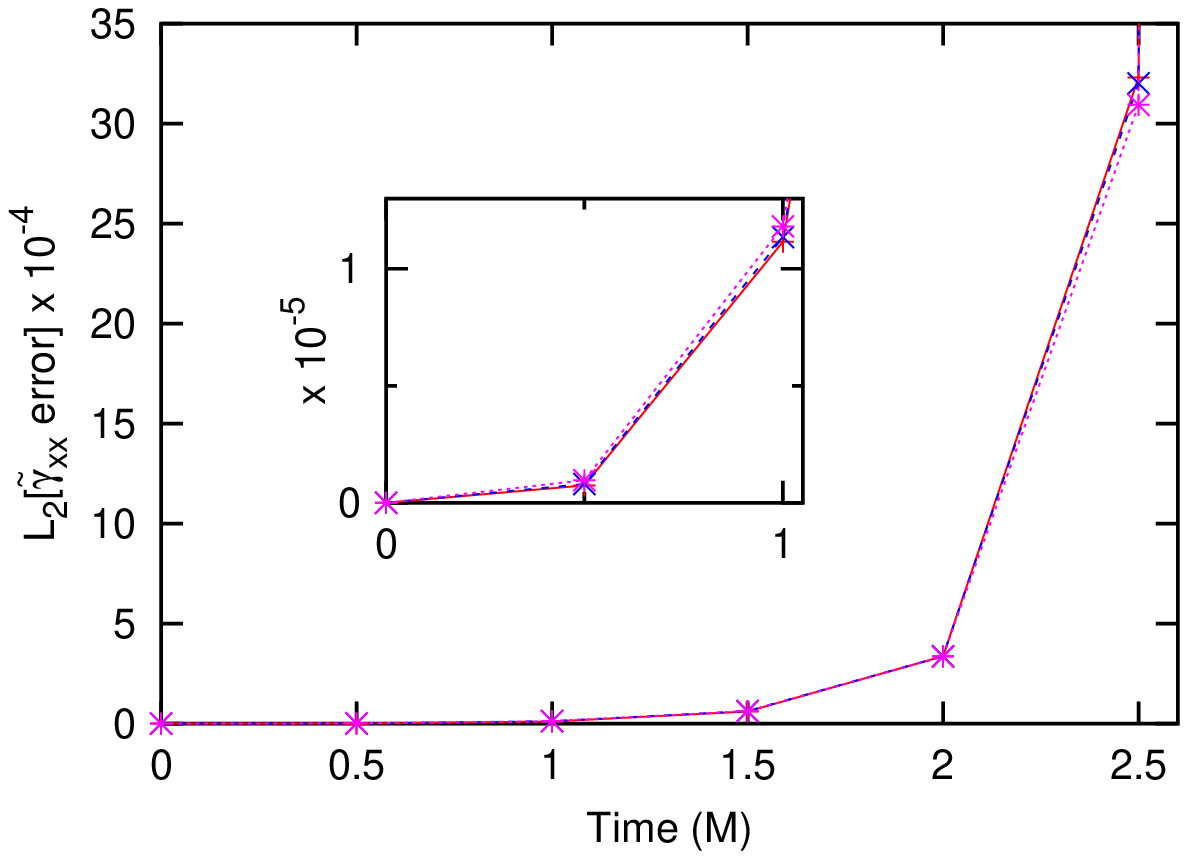}
\hspace*{\fill}
\\
\includegraphics[scale=.7,angle=-90]{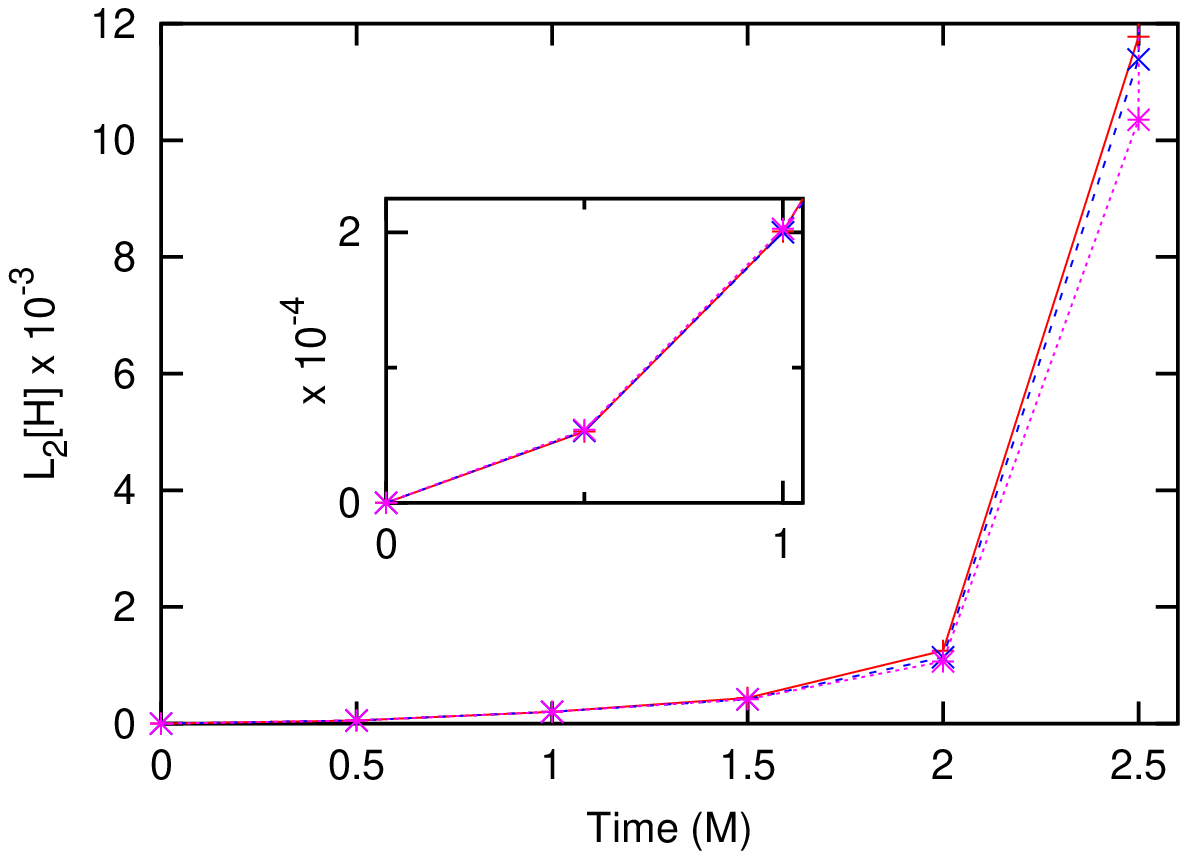}
\hspace*{\fill}
\caption{Convergence of the $L_2$ norms of the errors in $\tilde \gamma_{xx}$
and the Hamiltonian constraint $H$ for a geodesically sliced puncture.  The
numerical data points are here marked with the symbol ``x''. 
The solid lines show the errors for the highest
resolution run.  The errors for the medium resolution run (dashed lines)
and the lowest resolution run (dotted lines) have been divided by factors
of 4 and 16, respectively, to demonstrate second order convergence.}
\label{fig:norms_geod}
\end{figure}

\subsection{1 + log  slicing}
\label{sec:1+log_results}
Having rigorously tested the code against an analytic solution,
we now use a different coordinate condition
to study a longer-lived run with nontrivial, nonlinear dynamical
behavior in the region of FMR interface boundaries.
For this purpose, we again use zero shift but with a modified 
``1+log'' slicing condition given by 
\begin{equation}
\fpder{\alpha} {t} = -2\alpha \Psi_{BL}^4 K,
\label{eqn:1+log}
\end{equation}
where insertion of the factor
$\Psi_{BL} = 1 + {M}/{2r}$, 
originally recommended by 
\cite{Alcubierre02a}, has proven to enhance convergence
near the puncture in our simulations. 
For the numerical experiments with 1+log slicing, the grid structure, 
including
the locations of the mesh refinement interfaces, is the same as in the 
geodesic slicing case.  We carry out three runs, with low, medium
and high resolution defined as before.

A 1+log evolution serves as an excellent numerical experiment to test
the robustness of our mesh refinement interfaces.
The 1+log family has been well-studied in unigrid runs in the past, 
so the generic behavior of this coordinate system is known and provides
a general context for comparison with our mesh refinement results. 
 Because the 1+log slicing is singularity avoiding, in contrast
to the geodesic slicing case, simulations in a 1+log gauge are known to last
$\sim 30M - 40 M$, giving us an opportunity to study 
the properties of our mesh refinement interfaces in longer duration runs.
 Finally, as shown by
Fig.~\ref{fig:evol_1pluslog}, as the lapse (right panel of the figure)
\begin{figure*}
\includegraphics[scale=.7,angle=-90]{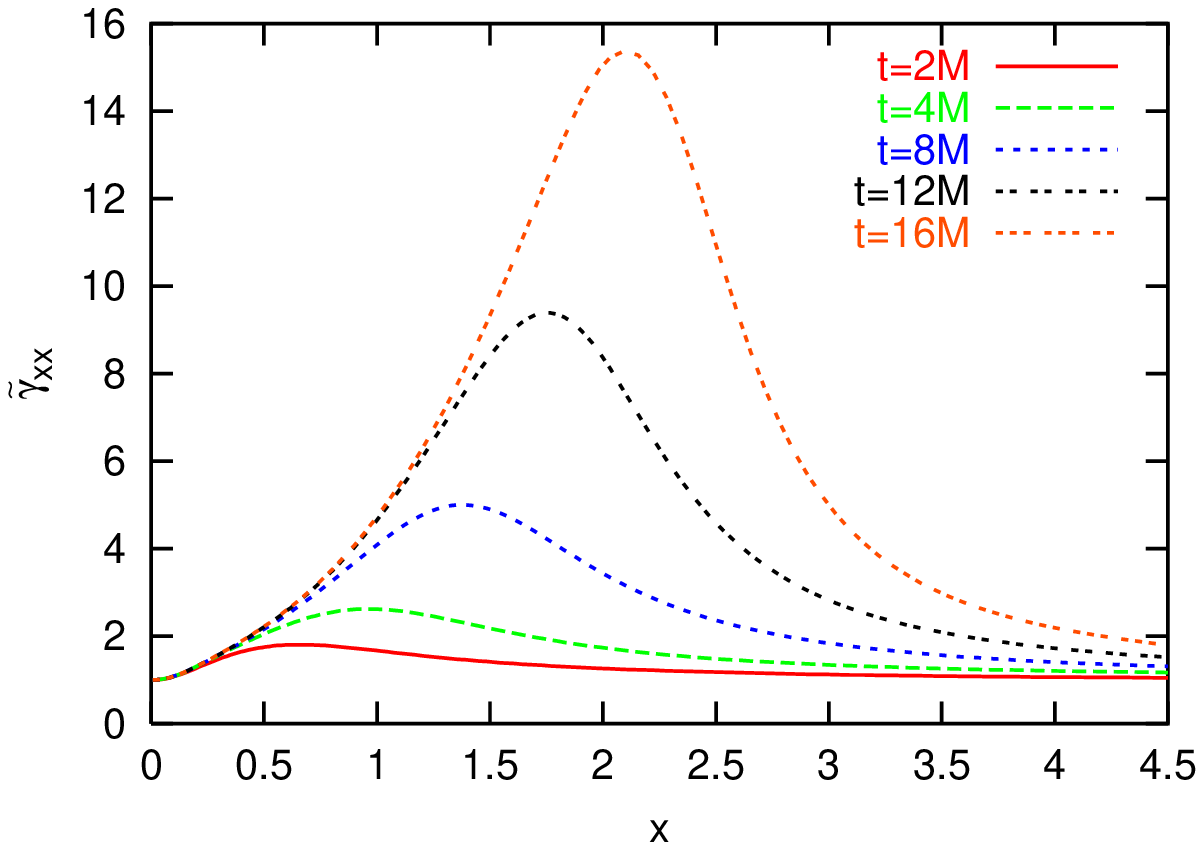}
\includegraphics[scale=.7,angle=-90]{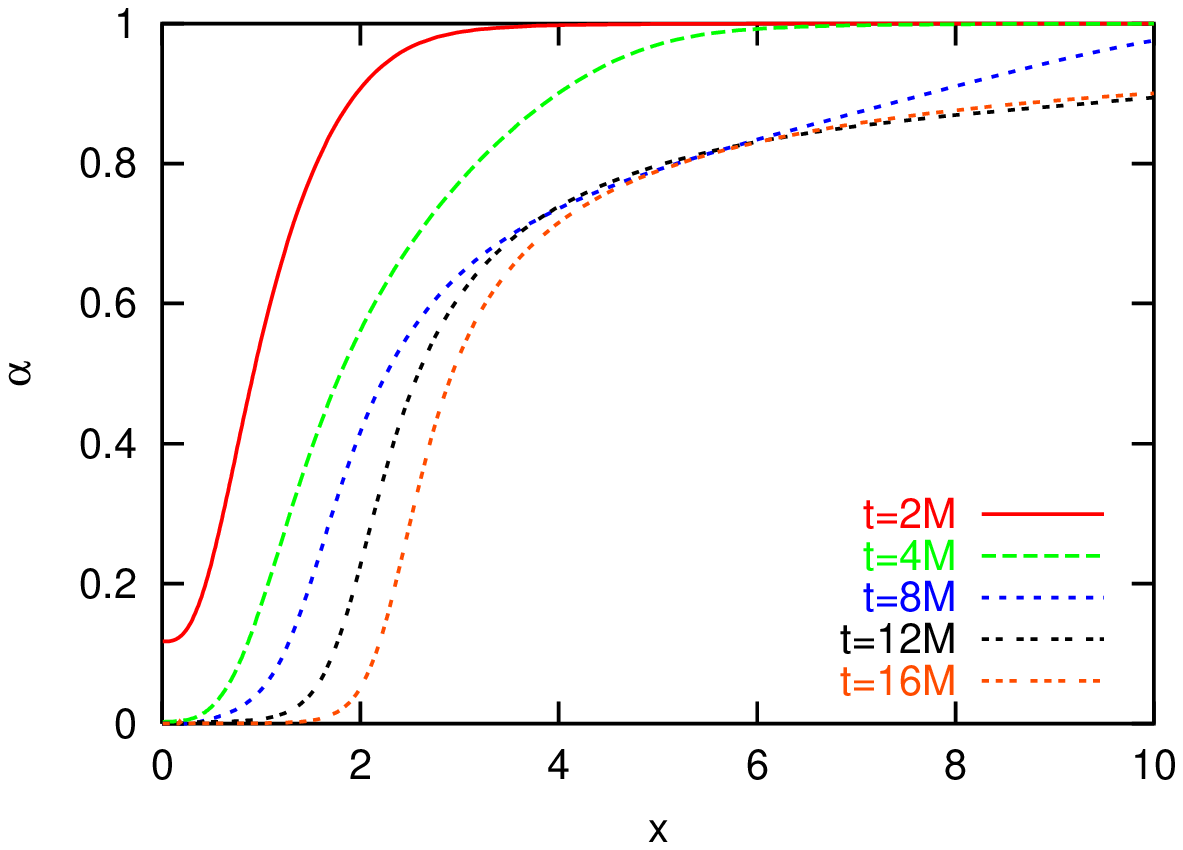}
\caption{A time evolution sequence for the conformal metric 
$\tilde{\gamma}_{xx}$ and the lapse $\alpha$ using the variant of
1+log slicing given in Eq.~(\ref{eqn:1+log}). Results are shown for the
highest resolution run.}
\label{fig:evol_1pluslog}
\end{figure*}
collapses around the singularity, a strong gradient region in the metric
(left panel of the figure) moves outward, passing through mesh refinement
boundaries in the process.  
According to unigrid runs already in the literature
({\emph {e.g.},} \cite{Alcubierre02a}) choosing an appropriate shift, such as the
Gamma-driver shift, would cause the evolution to freeze, preventing
catastrophic growth in the metric functions and confining the strong
field behavior to the region $r < 10M$.  This also increases the stable 
evolution time of the simulations.  
For our purposes here, however, we 
\emph{choose} to let the strong gradient region move outward because we
specifically wish to study how well the mesh refinement interfaces
handle a strong \emph{dynamical} potential on timescales $t > 10M$.
We consider this an important test, since such phenomena may develop
near refinement boundaries in the course of realistic astrophysical
simulations of multiple black holes.

Having made this choice, we expect to see exactly what appears in 
Fig.~\ref{fig:evol_1pluslog}.  The metric function
$\tilde\gamma_{xx}$ (left panel) grows due to well-understood
grid stretching related to the collapse of the lapse 
(right panel) and the fact that 
grid points are falling into the black hole.  The peak of the metric
simultaneously moves to larger coordinate position.  We expect, therefore,
that at some point certain regions of
the simulations will no longer exhibit second order convergence because
the gradients in the metric simply grow too large, because the peak of the
metric moves into a region of lower refinement that cannot resolve the
gradients already present in the metric at that point, or because of a 
combination of the two.  The simulations in this gauge, nonetheless, remain
second order convergent long enough for us to study the effects of the
strong potential passing through the innermost mesh refinement interfaces.

Because we do not have an analytic solution for the 1+log case to use in our 
convergence tests, we show three-point convergence plots instead.
Specifically, for a given field $f$, we plot
\((f_{\mathrm{low}} - f_{\mathrm{med}})/4\) using a dashed line
and
\(f_{\mathrm{med}} - f_{\mathrm{high}}\) using a solid line. Since 
the three different resolutions 
``low'', ``medium'', and ``high'' are related to each other by factors of two,
 the two lines in each panel should overlay exactly
for perfect second order convergence.

Fig.~\ref{fig:g+A_1pluslog} shows such a three-point convergence plot for
$\tilde{\gamma}_{xx}$ and $\tilde{A}_{xx}$ for a 1-D cut along the $x$-axis.
\begin{figure*}
\includegraphics[scale=.7,angle=-90]{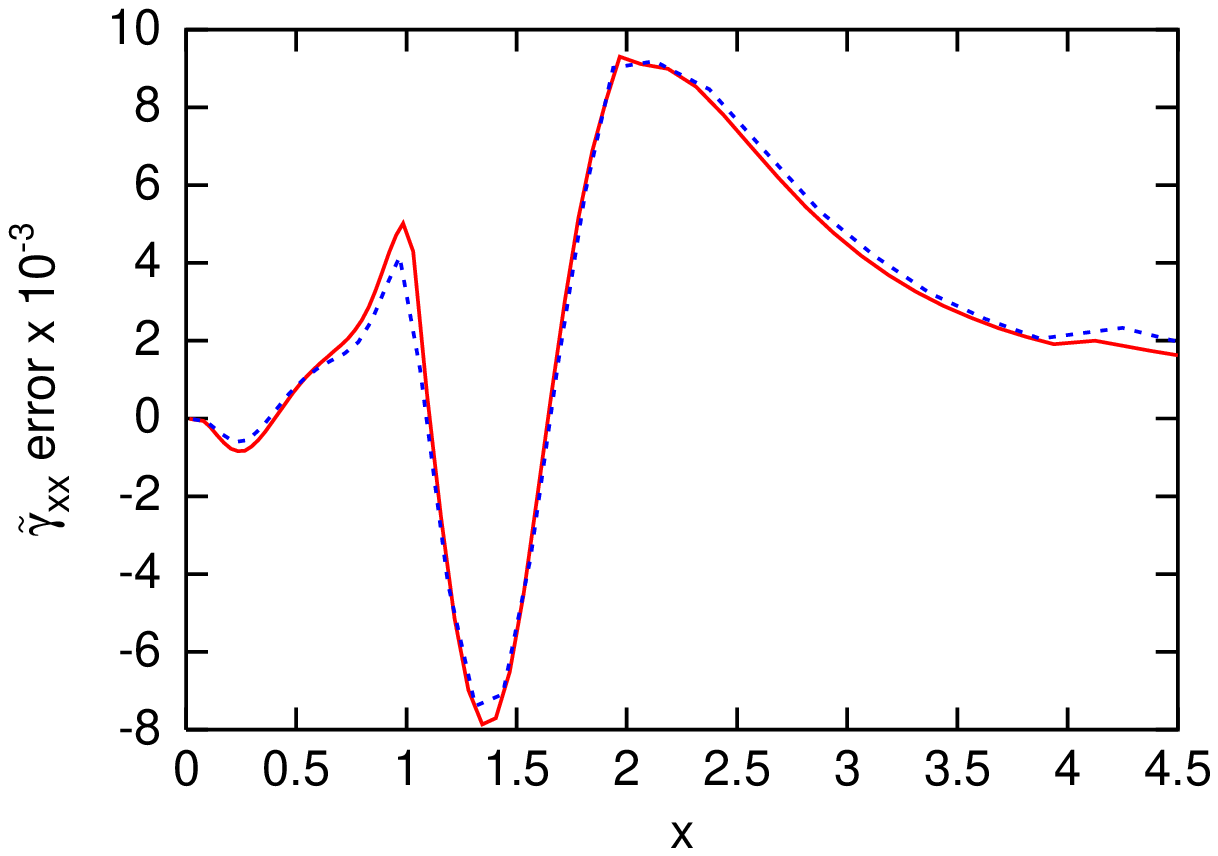}
\includegraphics[scale=.7,angle=-90]{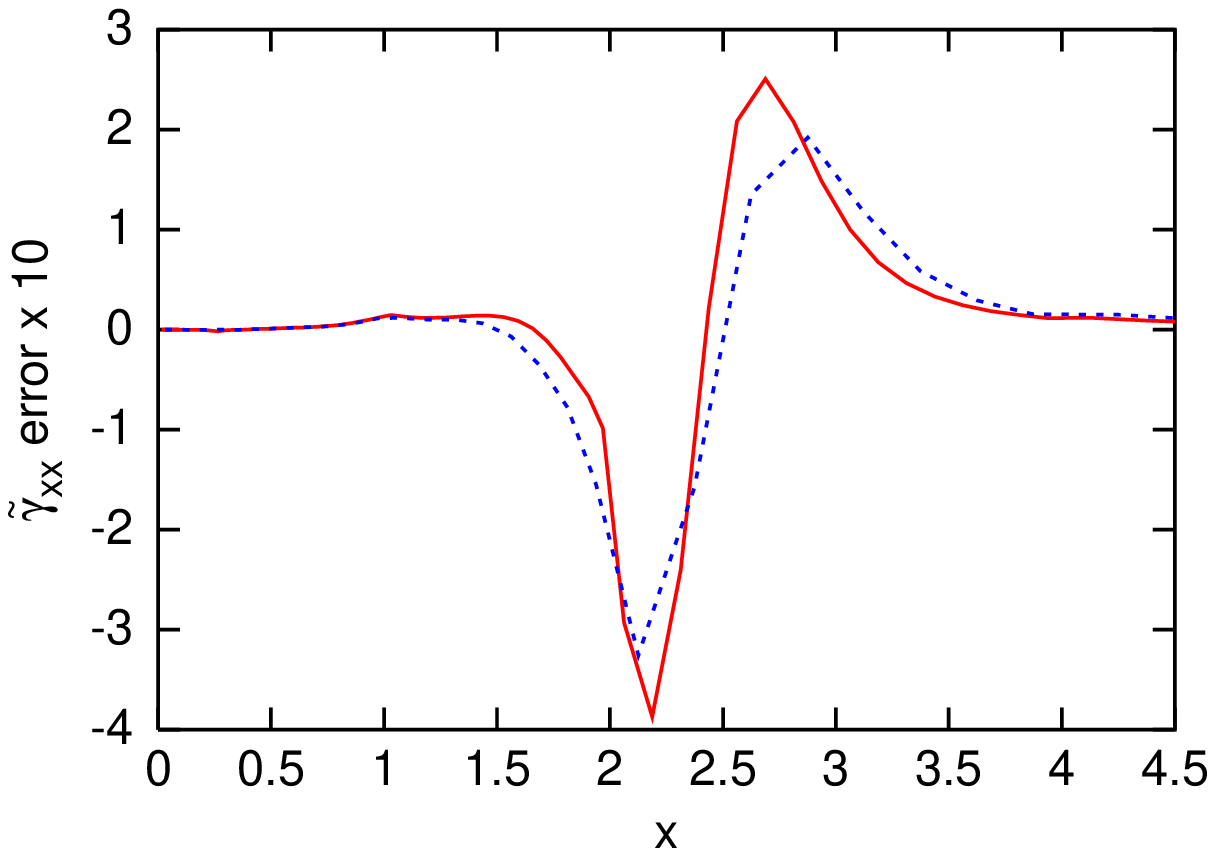}
\\
\includegraphics[scale=.7,angle=-90]{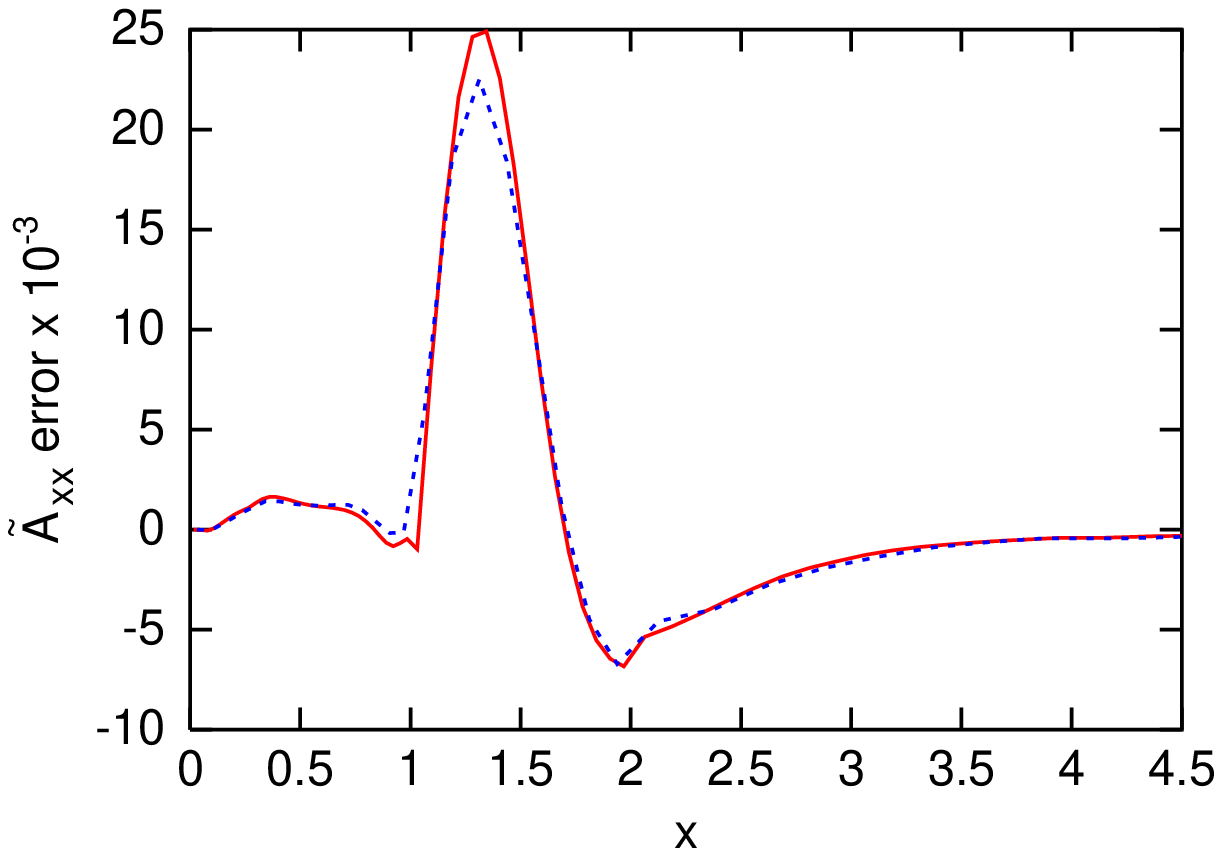}
\includegraphics[scale=.7,angle=-90]{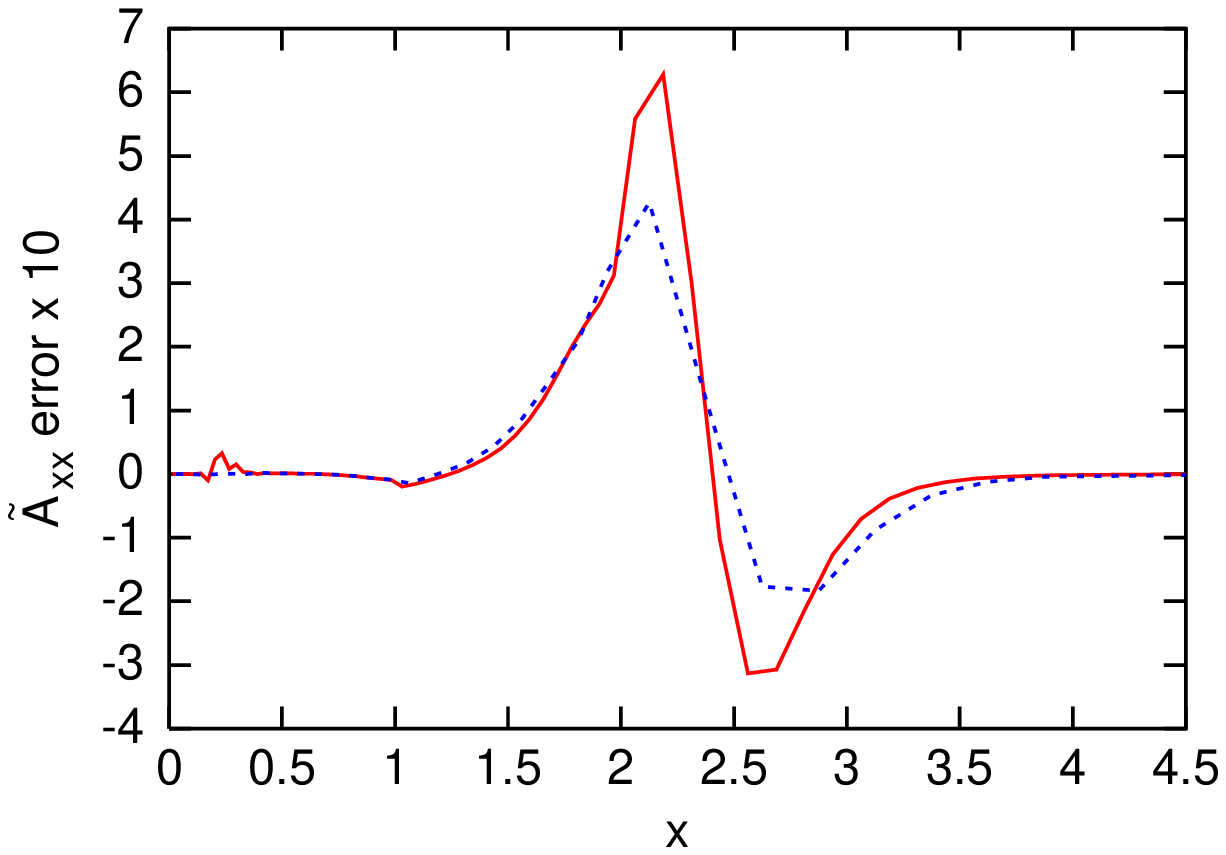}
\caption{Three-point convergence plots for the BSSN variables
$\tilde{\gamma}_{xx}$ and $\tilde{A}_{xx}$ along the $x$-axis in the 1+log slicing
runs at times $t=8M$ (left panels) and $t=16M$ (right panels). 
For a given field $f$, the dashed line
shows \((f_{\mathrm{low}} - f_{\mathrm{med}})/4\) and the solid line
shows \(f_{\mathrm{med}} - f_{\mathrm{high}}\).}
\label{fig:g+A_1pluslog}
\end{figure*}
The left panels, showing data from $t=8M$, demonstrate that the metric and 
other variables are second order convergent everywhere at that time.  
Overall, we continue to 
see second order convergence in the evolved variables, constraints, and
norms until  $t \sim 10M$. 

The convergent behavior starts to break down around $t \sim 10M$ due
to difficulties with resolving the sharp feature in the metric.  In 
the region $1M \leq x \leq 2M$, between the first and second FMR boundaries,
the peak itself grows sharply and the coarser grid is not sufficient
to provide the resolution needed for
convergent behavior.  For $2M \leq x \leq 4M$, the grid
is again coarsened by a factor of 2 and is not able to resolve 
adequately the steep gradient on the leading edge of the metric
peak. A snapshot at
$t=16M$ is shown in the right panels of Fig.~\ref{fig:g+A_1pluslog};
by this time, the peak of the metric has passed through
two refinement interfaces (at $x=1M$ and $x=2M$).
The time development of these errors, and in particular their departure
from second order convergence, can be seen in the animations available
in the EPAPS supplement; see Fig.\ 8A and the associated animation
file in Ref.\ \cite{EPAPS-FMR}.

Throughout the duration of the runs the region $x \gtrsim 5$ 
does remain second order convergent, even though the grid is
further coarsened by factors of two 
 at $x = 8M$, $16M$, $32M$, and $64M$, since
all the fields change very slowly as they approach the asymptotically
flat regime.
  The simulations will
continue to run stably past this point (to approximately $t \approx 35M$),
but 
the resolution in the regions to the right of the interface at $x=1M$ is
not sufficient to produce convergent results, as was expected.

The Hamiltonian and momentum constraints along the $x$-axis are shown in 
Fig.~\ref{fig:ham+mom_1pluslog}. Three curves are plotted in each
panel.  The errors for the highest resolution run are given by the solid
line.  The errors for the medium (dashed line) and low (dotted line)
resolution runs have been divided by factors of 4 and 16, respectively.
The constraints are second 
order convergent in the bulk for times $t \lesssim 10M$, when the resolution is
sufficient to handle the growing feature in the metric (left panels). 
As expected, $H$ exhibits first order convergent spikes at mesh refinement
interfaces; {\emph {cf.}}\ Fig.~\ref{fig:g-blowup} and Appendix~\ref{sec:errors}. 
\begin{figure*}
\includegraphics[scale=.7,angle=-90]{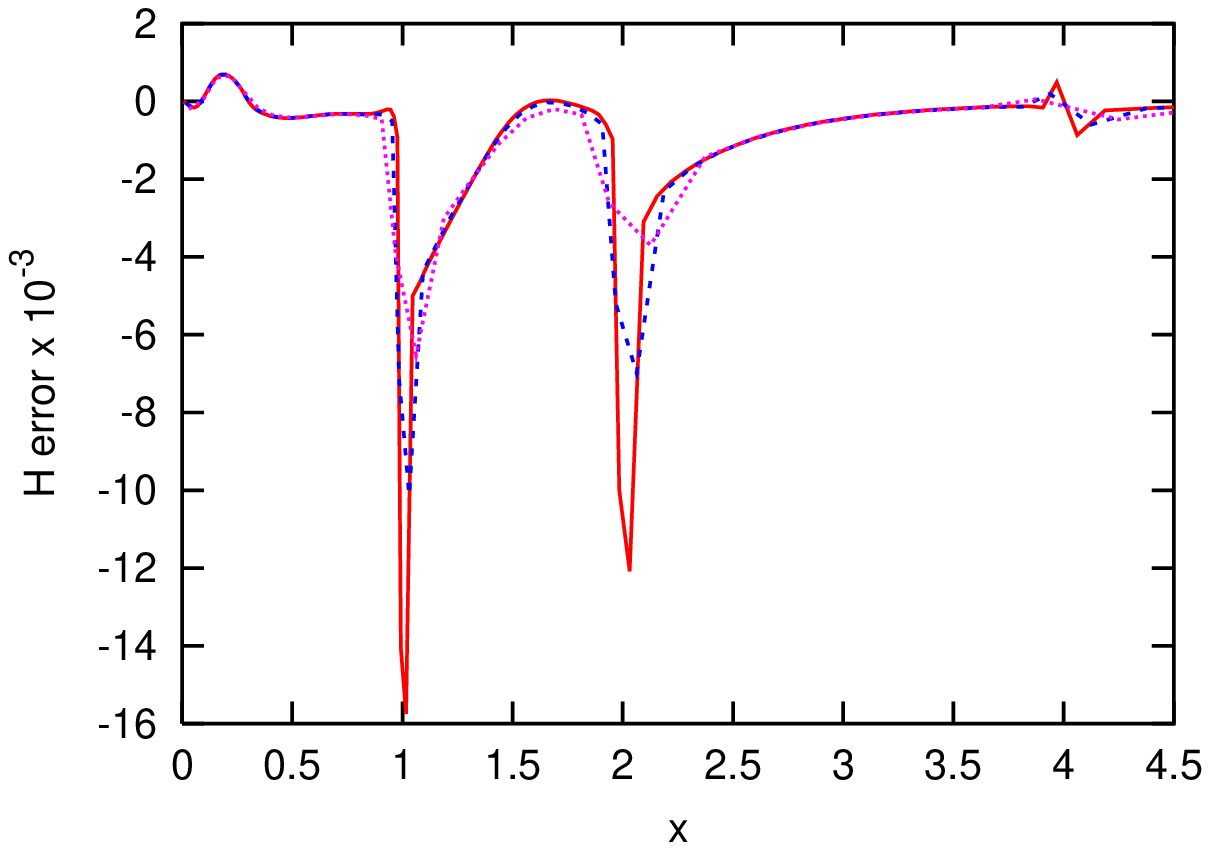}
\includegraphics[scale=.7,angle=-90]{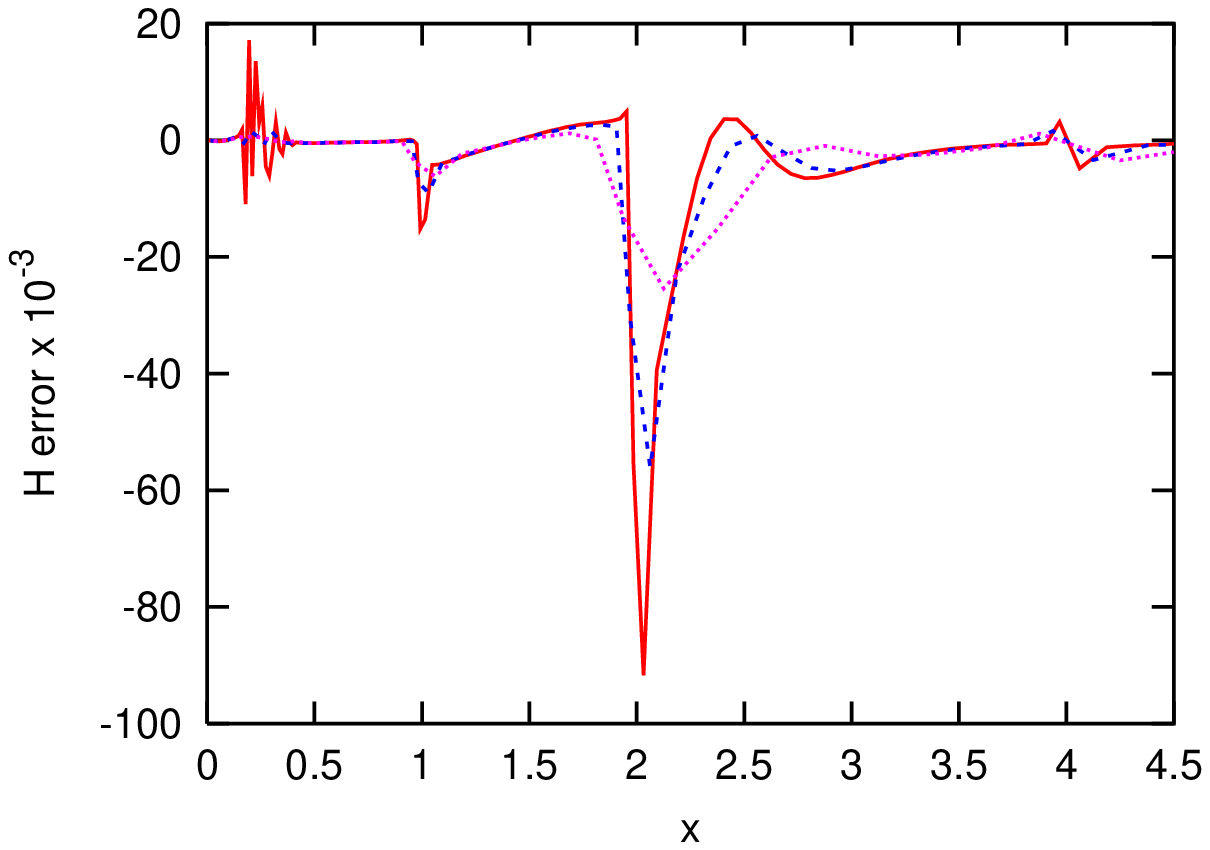}
\\
\includegraphics[scale=.7,angle=-90]{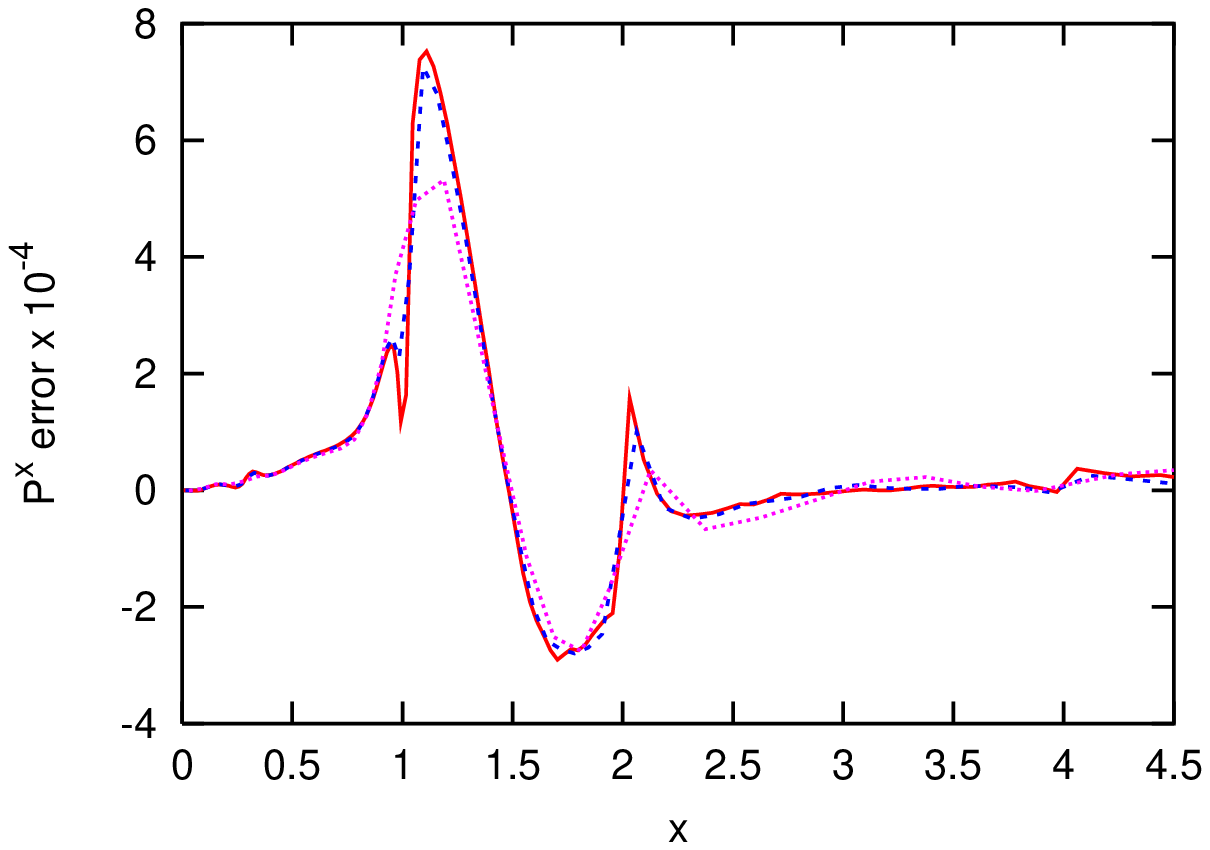}
\includegraphics[scale=.7,angle=-90]{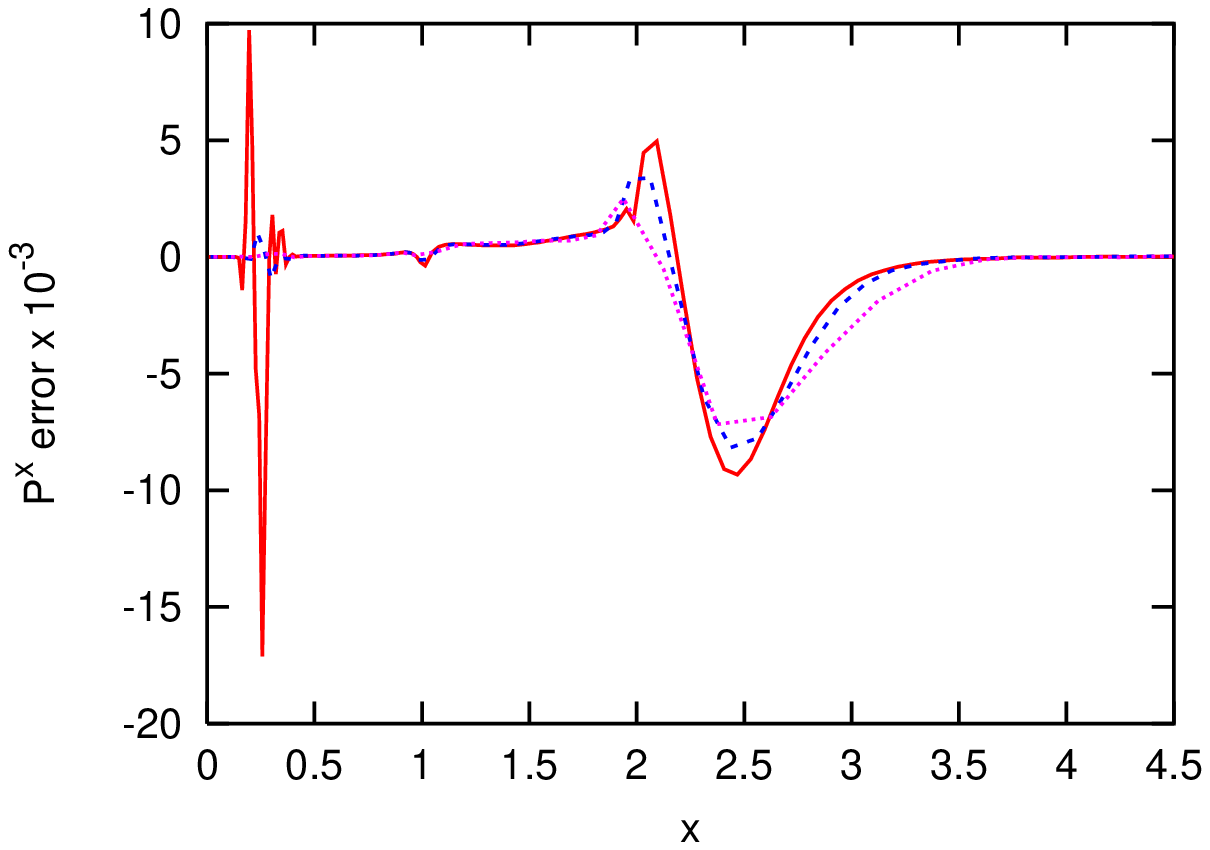}
\caption{Convergence plots for the Hamiltonian constraint $H$ and the
momentum constraint $P^{x}$ along the $x$-axis 
in the 1+log slicing runs at times $t=8M$ and $t=16M$.
The solid lines show the errors for the high resolution run.  The errors for
the medium (dashed line) and low (dotted line) resolution runs have been
divided by factors of 4 and 16, respectively.}
\label{fig:ham+mom_1pluslog}
\end{figure*}
For $t\gtrsim 10M$, as the peak of the fields propagates into the coarser 
grid regions past $x=2M$, the lowest
resolutions are not sufficient to resolve the rising slope of the metric,
and, like the evolved variables (Fig.~\ref{fig:g+A_1pluslog}), the constraints
no longer demonstrate second order convergence. 
The right panels of Fig.~\ref{fig:ham+mom_1pluslog} show the constraints
at $t=16M$, right after the peak of the metric passes through the refinement
interface at $x=2M$.   See Fig.\ 9A and the associated animation
file in Ref.\ \cite{EPAPS-FMR} 
for animations of these data.

The behavior of the simulations at locations away from the $x$-axis
is qualitatively similar to that shown in Figs.~\ref{fig:g+A_1pluslog}
and~\ref{fig:ham+mom_1pluslog}. Plots and animations of the errors along the
line $y = z = 0.25M$ are available in the EPAPS supplement;
see Figs.\ 8B and 9B and their associated animation files
in Ref.\ \cite{EPAPS-FMR}.

We have also examined the $L_1$ and $L_2$ norms of the errors to
assess the overall behavior of these runs, and display representative
results in Fig.~\ref{fig:1+log-norms}.  The $L_2$ norms
of the errors in the basic variables  
$\tilde \gamma_{xx}$  and $\tilde A_{xx}$ are shown in the left top and
bottom panels, respectively, using 3-point convergence plots. The
dashed lines show the difference between the low and medium resolution
results divided by 4, and the solid lines show the difference between
the medium and high resolution results, demonstrating the overall second
order convergence of these simulations at early times.
  The $L_1$ norm of $H$ is displayed in the top right panel,
where the solid line gives the errors for the
high resolution run. The errors for the medium (dashed line) and low
(dotted line) resolution runs have been divided by factors of 4 and
16, respectively to show second order convergence, as expected from
Eq.~(\ref{L1-boundaries+bulk}).  In the lower right panel the $L_2$ norm of $H$
is shown, 
with the solid line giving the results for the high resolution run.
As discussed in Appendix~\ref{sec:norms},
the errors for the medium (dashed line) and low (dotted line) resolution
runs have been divided by factors of $2^{3/2}$ and $4^{3/2} = 8$
to account for the effects of significant first order convergent
errors in $H$ at the mesh refinement boundaries, in addition to 
the second order convergent errors in the bulk; see 
Eq.~(\ref{L2-boundaries+bulk}).
 
One final feature of these simulations, the
high frequency noise  near the origin seen in the right panels of
Fig.~\ref{fig:ham+mom_1pluslog}, requires some explanation.
First of all, it is not related
to the presence of the refinement boundaries;
in particular, we have
reproduced it in unigrid runs and with an independently-written, 1-D 
(spherically symmetric) code. 
Higher resolution exacerbates this problem: both the frequency and 
amplitude of the noise increase with resolution.
 We have found the location of 
this noise to be independent of resolution and the number and positions of 
FMR boundaries. 

This feature, which we call the ``point-two $M$ problem,''
originates around $r \sim 0.2M$.
 It becomes most 
evident at times $t > 10M$, first appearing in the lapse and $K$, which 
are directly coupled, and then eventually mixing into all of the extrinsic 
curvature variables.  
For the duration of the evolutions the noise remains 
within the region $0.0M \lesssim x \lesssim 0.5M$. 
Outside this region, all basic variables demonstrate 
satisfactory second order convergence, including at refinement boundaries,
up to times $t \sim 10M$.

Having chosen a generally accepted gauge, and having focused on 
effects of the mesh refinement interfaces in this work, we have not
fully investigated the cause of nor possible remedies for this apparent
pathology.  We note it here, however, as an interesting topic for future
investigation.
\begin{figure*}
\hspace*{\fill}
\includegraphics[scale=.69,angle=-90]{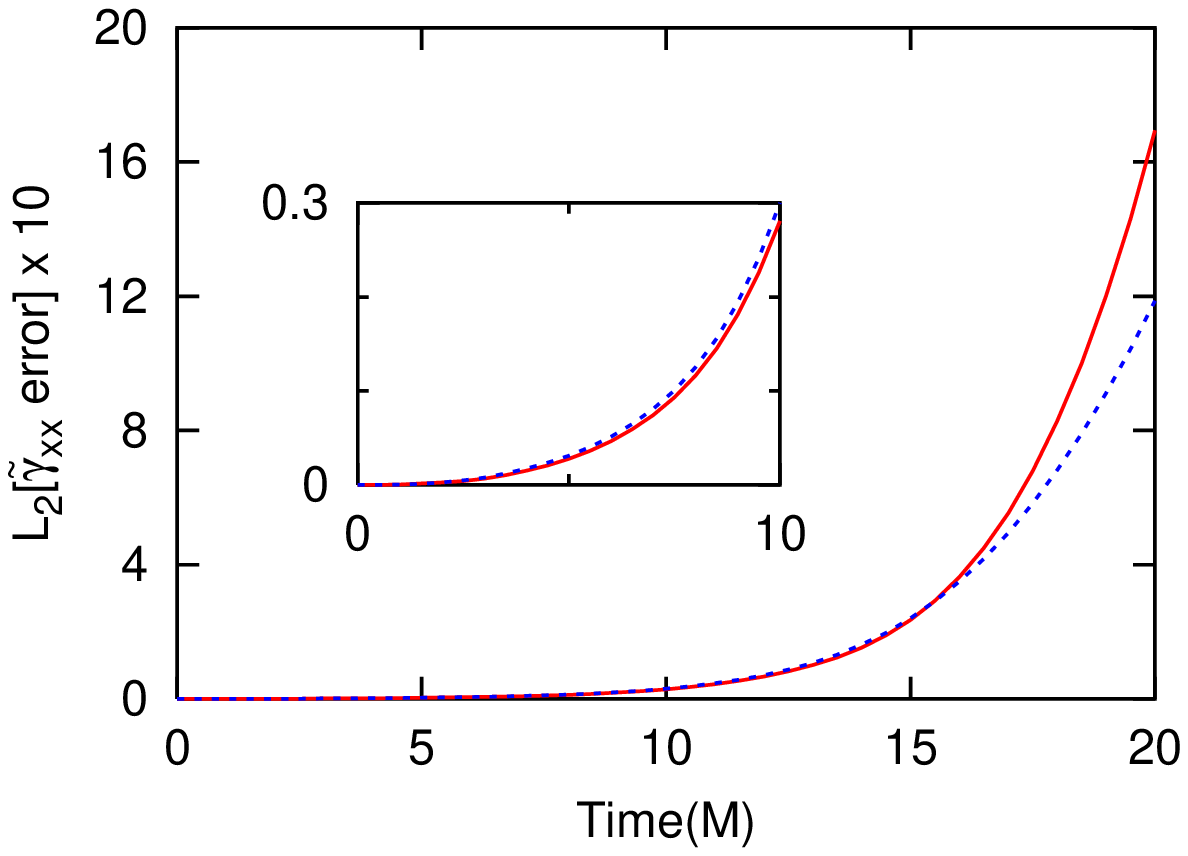}
\hspace*{\fill}
\includegraphics[scale=.69,angle=-90]{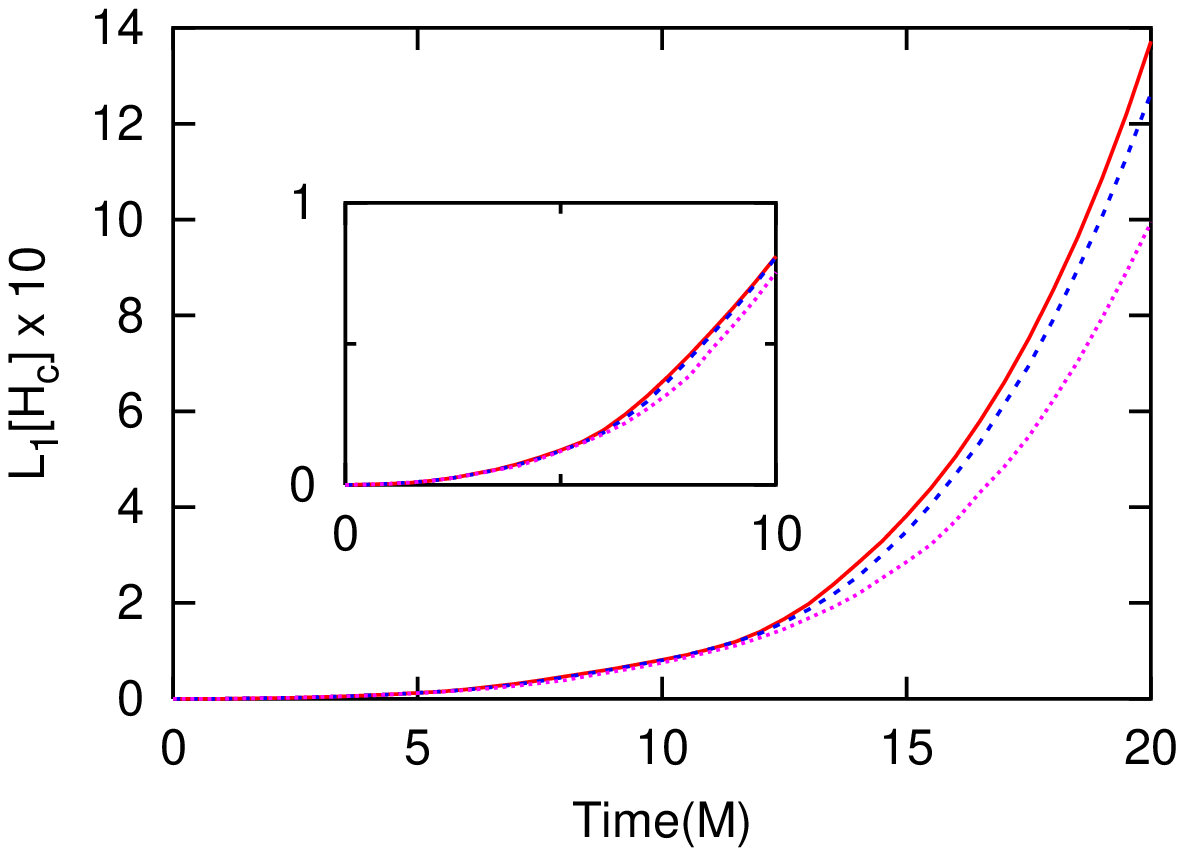}
\\
\hspace*{\fill}
\includegraphics[scale=.69,angle=-90]{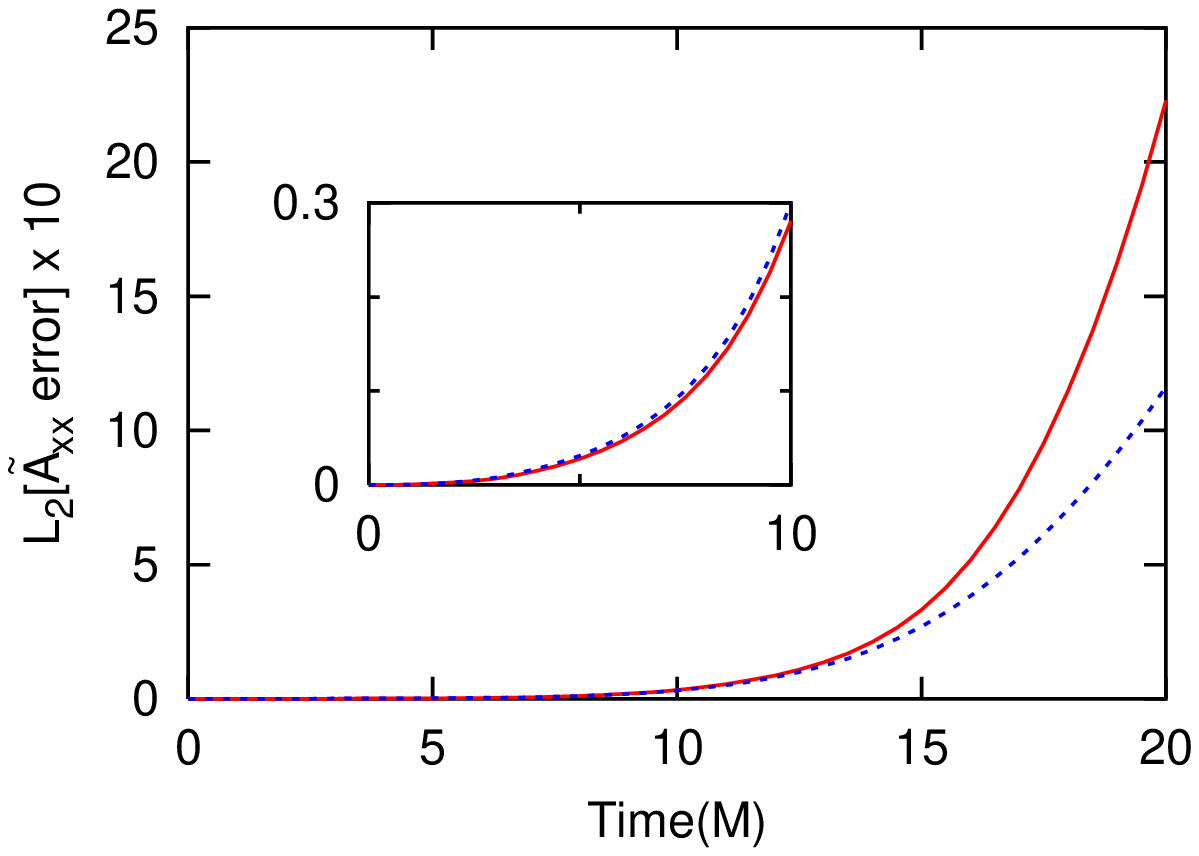}
\hspace*{\fill}
\includegraphics[scale=.69,angle=-90]{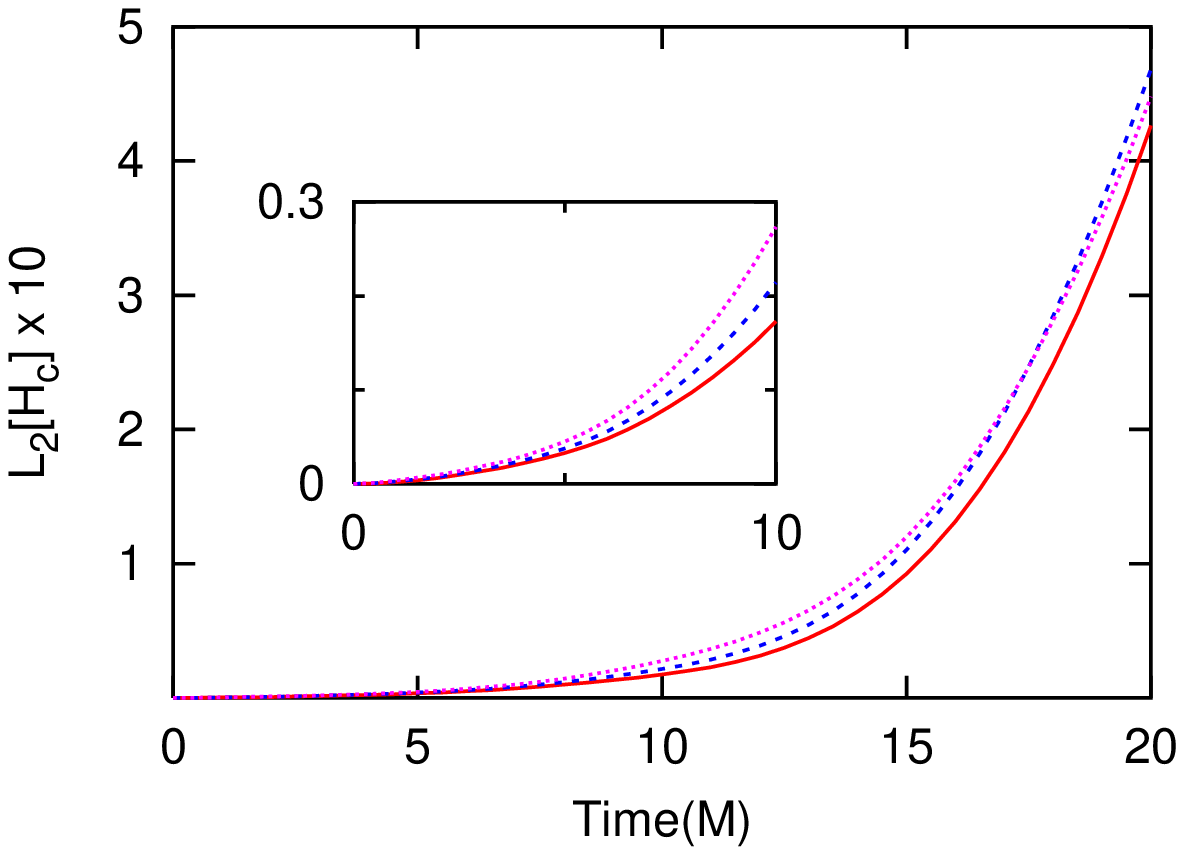}
\caption{Convergence behavior of the $L_1$ and $L_2$  norms of the errors
for the runs with 1 + log slicing.  The left panel shows the $L_2$ norms
of the errors in  
$\tilde \gamma_{xx}$ (top) and $\tilde A_{xx}$ (bottom), where the
dashed lines show the the difference between the low and medium resolution
results divided by 4, and the solid lines show the difference between
the medium and high resolution results.  The top right panel shows the
$L_1$ norm of $H$, with the solid line giving the errors for the
high resolution run; the errors for the medium (dashed line) and low
(dotted line) resolution runs have been divided by factors of 4 and
16, respectively.  The lower right panel shows the $L_2$ norm of $H$,
with the solid line giving the results for the high resolution run.
In this panel,
the errors for the medium (dashed line) and low (dotted line) resolution
runs have been divided by factors of $2^{3/2}$ and $4^{3/2} = 8$,
respectively.
}
\label{fig:1+log-norms}
\end{figure*}

\section{Summary}
\label{sec:summary}
This paper demonstrates that fixed mesh refinement boundaries can be located
in the \emph{strong field region} of a \emph{dynamical} black hole spacetime 
when the interface conditions are handled properly. 
This result was verified through 
simulation of a Schwarzschild black hole in geodesic coordinates, for which
we have an analytic solution for comparision, and through simulations of
Schwarzschild in a variation of the 1+log (singularity avoiding) slicing
with zero shift.
Mesh refinement technology, therefore, 
is a viable to way to use computational resources
more efficiently, and to simulate the very large spatial domains
needed to compute the dynamics of the source interactions and 
allow extraction of the resulting gravitational waveforms.  

Our method for handling the interface
conditions, based in part on the Paramesh infrastructure, is detailed.
For these simulations
we find that, in handling the interface condition between FMR levels, 
third order guard cell filling is sufficient for overall second
order accuracy in the simulations.
By nesting several levels of mesh refinement regions, we are able to 
resolve the puncture convergently while simultaneously pushing the 
outer boundary of 
our domain to $128M$ 
and keeping the computational problem size modest.  
We estimate that for only a 12\% increase in the computational size of
the problem, we could push the outer boundary to $256M$; moving the
outer boundary out even farther will be possible for production runs on
larger machines. 
 Combined with our
earlier results showing that gravitational waves pass through such FMR 
interfaces without significant reflections \cite{Choi:2003ba}, we have now
studied, in detail, the effects of FMR interfaces on the two primary features,
waves and time-varying strong potentials, of astrophysically interesting 
spacetimes.

In this paper, we have evolved single black holes
using gauges with zero shift in order to produce test
problems in which strong-field spacetime features with steep gradients
pass through mesh refinement interfaces. 
In more realistic, astrophysical simulations of multiple black holes,
we expect to use non-zero shift prescriptions.
While a shift vector will allow us
to control certain aspects of the dynamics, we still expect to find
some strong, time-varying signals to propagate across mesh refinement
boundaries.
We are currently implementing non-zero shift conditions into our FMR
evolutions and will report on this work in a separate publication.

\begin{acknowledgments}
It is a pleasure to thank Richard Matzner for his penetrating
discussions about convergence of code results, and 
 Peter MacNeice for his enlightening replies to our queries about
Paramesh and other code related issues.  
We are also grateful to Josephine Palencia and Jeff Simpson,
who provided essential computer support during the course of this 
investigation.  This work was performed while authors J.B. and J.v.M. held 
National Research Council Associateship Awards at the 
Goddard Space Flight Center, and was 
funded by NASA Space Sciences grant ATP02-0043-0056. JDB 
also received funding from NSF grant PHY--0070892. The computations were
carried out on Beowulf clusters operated by the Space Science Data
Operations Office and the Commodity Cluster Computing Group at
Goddard. 
\end{acknowledgments}

\appendix

\section{Error Analysis of Guard Cell Filling Scheme}
\label{sec:errors}

To help pave the way for understanding the behavior of our black hole 
evolutions near mesh refinement boundaries, we provide here a detailed
analysis of  
a toy model for a scalar field in one spatial dimension,
using the same third order guard cell filling algorithm detailed in
Sec.~\ref{sec:interfaces}. The model equations are
\begin{subequations} \label{eqn:SF}
\begin{eqnarray}
  \dot\psi & = &  \pi \   \label{eqn:SF_a} \\
  \dot\pi  & = &  \psi'' \ ,\label{eqn:SF_b}
\end{eqnarray}
\end{subequations}
where the dot denotes a time derivative and primes 
denote space derivatives. These equations can be solved numerically using the 
same twice--iterated 
Crank--Nicholson algorithm used to evolve our black hole
spacetimes. The fields at timestep $n+1$ 
are given in terms of the fields at timestep $n$ by
\begin{subequations} \label{eqn:discreteSF}
\begin{eqnarray}
  \psi^{n+1}_j & = & \psi^n_j + {\Delta t}\, \pi^n_j + \frac{\Delta t^2}{2} D^2\psi^n_j
  + \frac{\Delta t^3}{4} D^2\pi^n_j \   \label{eqn:discreteSF_a}\\
  \pi^{n+1}_j & = & \pi^n_j + {\Delta t}\, D^2\psi^n_j + \frac{\Delta t^2}{2} D^2\pi^n_j 
  + \frac{\Delta t^3}{4} D^4\psi^n_j \quad  \label{eqn:discreteSF_b}
\end{eqnarray}
\end{subequations}
where $D^2$ is a finite difference operator approximating the second spatial 
derivative, and $D^4 = (D^2)^2$. 

Consider for the moment a uniform spatial grid. If $D^2$ is the usual second order 
accurate centered difference operator, the 
dominant source of error for $\psi^{n+1}_j$ comes from the term proportional to 
$\Delta t^3$. This term has the wrong numerical coefficient as compared to the Taylor series expansion of the 
exact solution. The dominant sources of error for $\pi^{n+1}_j$ come from the term proportional to 
$\Delta t^3$, which also has the wrong numerical coefficient, and from the second order error in 
$D^2\psi^n_j$. For a uniform grid the dominant error in $D^2\psi^n_j$ is 
$D^4\psi^n_j\, \Delta x^2/12$, so the leading errors for a single timestep are
\begin{subequations} \label{eqn:SFerror}
\begin{eqnarray}
  (\psi^{n+1}_j)_{err} & = & \frac{1}{12} \Delta t^3 D^2\pi^n_j \  \label{eqn:SFerror_a}\\
  (\pi^{n+1}_j)_{err}  & = & \frac{1}{12} \Delta t\,(\Delta t^2 + \Delta x^2) D^4\psi^n_j 
    \ .\label{eqn:SFerror_b}
\end{eqnarray}
\end{subequations}
For $\Delta t \sim \Delta x$, each of these one--time--step errors is proportional to $\Delta x^3$. 
If we evolve the initial data to a finite time $T$, the ${\cal O}(\Delta x^3)$ errors accumulate over 
$N = T/\Delta t$ timesteps resulting in second order errors.\footnote{This is a 
simplification. The dominant error after $N$ timesteps includes other terms of order $\Delta x^2$ 
in addition to the product of $N$ and the one--time--step error. These other terms include, 
for example, the product of an order $N^3$ coefficient 
and a one--time--step error of order $\Delta x^5$.} Thus, the basic variables 
$\psi$ and $\pi$ are second order convergent on a uniform grid.

On a non--uniform grid, guard cell filling introduces errors of order $\Delta x^3$ in $\psi$ at grid 
points adjacent to the boundary. This leads to errors of order $\Delta x$ in $D^2\psi^n_j$ and  
$1/\Delta x$ in $D^4\psi^n_j$. From Eq.~(\ref{eqn:discreteSF_b}) we see that in one timestep 
$\pi$ can acquire errors of order $\Delta x^2$. The concern is that these errors might accumulate over 
$N = T/\Delta t$ timesteps to yield first order errors. This, in fact, does not happen. Simple numerical 
experiments show that $\psi$ and $\pi$ are second order convergent on a non--uniform grid with 
third order guard cell filling. 

We can understand this result with the following heuristic 
reasoning. The numerical algorithm of Eq.~(\ref{eqn:discreteSF}) 
approximates, as does any mathematically sound numerical scheme,
 the exact solution of the scalar field equations 
(Eq.~(\ref{eqn:SF})) in which the field $\pi$ propagates along the light cone. 
The ``bulk'' errors displayed in 
Eq.~(\ref{eqn:SFerror_b}) accumulate along the past light cone to produce an overall error of order 
$N\,\Delta x^3 \sim \Delta x^2$ at 
each spacetime point. Errors in guard cell filling, which 
occur at a fixed spatial location, do not accumulate over multiple timesteps 
since the past light cone of a given 
spacetime point will cross the interface (typically) no more than once.

The characteristic fields for the system (\ref{eqn:SF}) are $\pi \pm \psi'$ so that 
$\psi'$, like $\pi$, propagates along the light cone. As a result, the value of $\psi$ 
at a given spacetime point is determined by data from the {\it interior} of the past light cone. 
From Eq.~({\ref{eqn:discreteSF_a}}) we see that the 
one--time--step errors for $\psi$ due to guard 
cell filling are order $\Delta x^3$. These errors can accumulate over $N$ timesteps to yield errors 
of order $N\, \Delta x^3\sim \Delta x^2$.

The derivatives $\psi'$ and $\psi''$  are computed at finite time $T$ by evolving the 
$\psi$, $\pi$ system for $T/\Delta t$ timesteps then taking the centered, second order accurate 
numerical derivatives of $\psi$. Numerical experiments show that $\psi'$ and $\psi''$, defined in 
this way, are second order convergent on a non--uniform grid with third order guard cell 
filling.\footnote{The error in $\psi''$ 
is fairly noisy but the overall envelope containing this noise is second order convergent.} 
Continuing with our heuristic discussion, we can understand the second order convergence of 
$\psi''$ as follows. Let $(\psi^n_{j})_{err} \approx E^n_j \Delta x^2$ denote the error
 in $\psi$ at grid point $n$, $j$, where the coefficient $E^n_j$ is 
independent of $\Delta x$. Some of this error 
is due to guard cell filling at the mesh refinement 
interface and some is due to the accumulation of ``bulk'' errors 
(\ref{eqn:SFerror}). Now, the second derivative of $\psi$, computed as 
$D^2\psi^n_j = (\psi^n_{j+1} - 2\psi^n_j + \psi^n_{j-1})/{\Delta x}^2$, will
contain errors of the form $(D^2\psi^n_j)_{err} = E^n_{j+1} 
- 2E^n_j + E^n_{j-1}$. Since the bulk errors are smooth, the bulk contribution to 
$E^n_{j+1} - 2E^n_j + E^n_{j-1}$ will scale as $\Delta x^2$. It  is also the case that 
the errors due to guard cell filling are smooth. This is because the value of $\psi$ at any 
given point is 
determined by the interior of the past light cone, so its error includes an accumulation of 
guard cell filling errors along the history of the mesh refinement interface. In the limit of high resolution 
this accumulation of error approaches the same value at neighboring grid points $j-1$, $j$, and $j+1$. 
In other words, the guard cell filling contribution to $E^n_{j+1} - 2E^n_j + E^n_{j-1}$ approaches 
zero as $\Delta x \to 0$. Evidently, the guard cell filling contribution, 
like the bulk contribution, scales like  $\Delta x^2$. 

The discussion above indicates that we can evolve the scalar field system of Eq.~({\ref{eqn:SF}}) 
for a finite time on a grid with mesh refinement, numerically compute the second derivative of 
$\psi$, and find that 
$\psi''$ is second order accurate. Without shift, the BSSN 
equations (\ref{eqn:gevol}) and (\ref{eqn:Aevol}) are similar to the scalar 
field equations with $\tilde\gamma_{ij}$ playing the role of $\psi$ and $-\tilde A_{ij}$ 
playing the role of $\pi$. This feature was one of the original motivations behind the 
BSSN system. Note that the term analogous to $\psi''$ in the $\dot\pi$ equation is the term 
$\tilde\gamma^{l m} \tilde\gamma_{i j,l m}$ contained in the trace--free part of the Ricci tensor, 
which appears on the right--hand side of Eq.~(\ref{eqn:Aevol}). Obviously there are many other terms 
that appear on the right--hand side of the $d\tilde A_{ij}/dt$ equation. We can model the effect of 
these terms by including a fixed function on the right--hand side of the $\dot\pi$ equation:
\begin{subequations} \label{eqn:SFplus}
\begin{eqnarray}
  \dot\psi & = & \pi \  \label{eqn:SFplus_a}\\
  \dot\pi  & = & \psi'' - \chi'' \ .\label{eqn:SFplus_b}
\end{eqnarray}
\end{subequations}
We have written the fixed function as the second derivative of $\chi$.  For simplicity we 
choose $\chi$ to depend on $x$ only, the most relevant dependence for our consideration of behavior
across spatial resolution interfaces.   The general solution of this system is then
\begin{subequations} \label{eqn:SFplussolution}
\begin{eqnarray}
  \psi(t,x) & = & \bar\psi(t,x) + \chi(x) \  \label{eqn:SFplussolution_a}\\
  \pi(t,x)  & = & \bar\pi(t,x) \  \label{eqn:SFplussolution_b}
\end{eqnarray}
\end{subequations}
where $\bar\psi$, $\bar\pi$ is a solution of the homogeneous wave equation (Eq.~(\ref{eqn:SF})). 

The extended model system of Eq.~(\ref{eqn:SFplus}) can be solved numerically with the discretization 
\begin{subequations} \label{eqn:discreteSFplus}
\begin{eqnarray}
  \psi^{n+1}_j & = & \psi^n_j + {\Delta t}\, \pi^n_j + \ldots \   \label{eqn:discreteSFplus_a}\\
  \pi^{n+1}_j & = & \pi^n_j + {\Delta t}\, (D^2\psi^n_j - D^2\chi_j) + \ldots 
  \  \label{eqn:discreteSFplus_b}
\end{eqnarray}
\end{subequations}
The higher order terms in $\Delta t$, not shown here, come from the iterations in our iterated 
Crank--Nicholson algorithm. 
It is important to recognize that the  $\chi''$ term is expressed as the numerical second derivative
of $\chi_j$ and not as the discretization of the 
analytical second derivative, $(\chi'')_j$. The reason for this choice is that $D^2\chi_j$ 
mimics the effect of the extra terms on the right--hand side of Eq.~(\ref{eqn:Aevol}) which, 
in our BSSN code,  depend on the discrete first and second derivatives of the BSSN 
variables $\phi$ and $\tilde\Gamma^i$. 

From the discussion of the wave equation (Eq.~(\ref{eqn:SF})) we can anticipate the results of 
numerical experiments with the model system (Eq.~(\ref{eqn:SFplus})) on a non--uniform grid.  
For arbitrary initial data 
$\psi^0_j$, $\pi^0_j$, the numerical solution is given by 
\begin{subequations}\label{eqn:SFdiscsoln}
\begin{eqnarray}
  \psi^n_j & = & \bar\psi^n_j + \chi_j \  \label{eqn:SFdiscsoln_a}\\
  \pi^n_j & = & \bar\pi^n_j \  \label{eqn:SFdiscsoln_b}
\end{eqnarray}
\end{subequations}
where $\bar\psi^n_j$, $\bar\pi^n_j$ is the numerical solution of the 
homogeneous wave equation (Eq.~(\ref{eqn:SF})) with initial data 
$\bar\psi^0_j - \chi_j$, $\bar\pi^0_j$. The order of convergence for $\psi^n_j$ is determined by 
how rapidly, as $\Delta x \to 0$, the numerical solution in Eq.~(\ref{eqn:SFdiscsoln_a})
approaches the exact solution $\psi(t,x) = \bar\psi(t,x) + \chi(x)$. 
Since $\chi_j$ is simply the projection of the analytic function $\chi(x)$ onto the numerical 
grid, the term $\chi_j$ in the numerical solution (Eq.~(\ref{eqn:SFdiscsoln_a})) does not contribute any error.
We have already determined 
that on a non--uniform grid $\bar\psi^n_j$ approaches $\bar\psi(t,x)$ with second order accuracy. 
Thus, we expect $\psi^n_j$ to be second order convergent.

What about derivatives of $\psi$? The order of convergence for $D^1\psi^n_j$ is found by comparing 
the discrete derivative $D^1\psi^n_j = D^1\bar\psi^n_j + D^1\chi_j$ to the analytic solution 
$\psi' = \bar\psi' + \chi'$. Again, as we have discussed, $D^1\bar\psi^n_j$ approaches $\bar\psi'$ with 
second order errors. It is also easy to see that the numerical derivative $D^1\chi_j$ 
approaches $\chi'$ with second order 
accuracy. Away from grid interfaces this is obviously true, assuming that $D^1$ is the 
standard second order accurate centered difference operator. 
For points adjacent to a grid interface, guard cell 
values for $\psi$ are filled with third order errors. These errors lead to second order errors in 
$D^1\psi^n_j$. Overall then, we expect second order convergence for $D^1\psi^n_j$. 

The expected convergence rates for $\psi$ and $\psi'$ are confirmed by the results shown in 
Fig.~\ref{psifigure}. For these numerical tests, we chose $\chi(x) = \exp((x-50)/10)$ and
initial data 
\begin{subequations} \label{eqn:SFindat}
\begin{eqnarray}
  \psi(0,x) & = & 100 e^{-(x+10)^2/400} + e^{(x-50)/10} \  \label{eqn:SFindat_a}\\ 
  \pi(0,x) & = & \frac{1}{2} (x+10) e^{-(x+10)^2/400} \ .\label{eqn:SFindat_b}
\end{eqnarray}
\end{subequations}
Each set of curves shows the errors at three 
different resolutions, $\Delta x = 5/16$, $5/32$, and $5/64$, where $\Delta x$ is 
the fine grid spacing. The evolution time is $20.83$, corresponding to $200$, $400$, or 
$800$ timesteps (depending on the resolution) and a Courant factor of $1/3$.
\begin{figure}
\includegraphics[scale=.7,angle=-90]{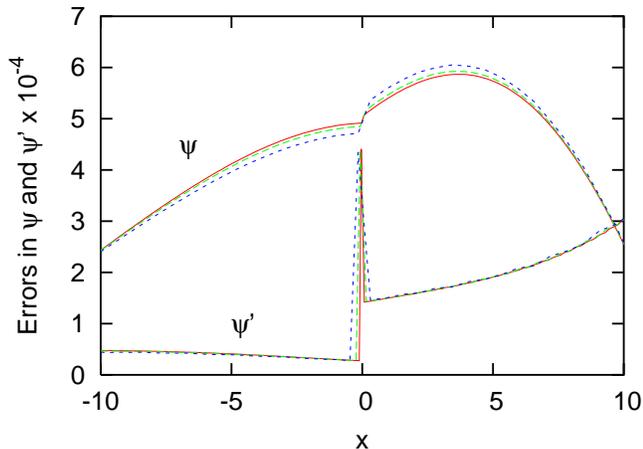}
\caption{Convergence tests for $\psi$ and $\psi'$. The mesh interface is at 
$x=0$, with the fine grid on the left and 
coarse grid on the right. The errors in $\psi$ and $\psi'$ for the high
resolution case are shown by the solid line.  The errors are
divided by 
factors of 4 and 16 for the middle (dashed line) and low 
(dotted line) resolution cases, respectively.}
\label{psifigure}
\end{figure}

The order of convergence for the second derivative of $\psi$ is determined from a comparison of 
$D^2\psi^n_j = D^2\bar\psi^n_j + D^2\chi_j$ and the analytic solution $\psi'' = \bar\psi'' + \chi''$. 
We have seen that 
$D^2\bar\psi^n_j$ approaches $\bar\psi''$ with second order accuracy. The situation for $D^2\chi_j$, however, 
is somewhat different. Away from any grid interface $D^2\chi_j$ will approach $\chi''$ with second 
order accuracy, assuming $D^2$ is the standard second order accurate finite difference operator. But for 
points adjacent to the interface, and only those points, guard cell filling errors of order 
$\Delta x^3$ in $\chi$ will lead to first order errors in $D^2\chi_j$. Thus, we expect to find 
second order convergence for $D^2\psi^n_j$ at all points except those points adjacent to the interface. Points 
adjacent to the interface should be first order convergent. 

\begin{figure}
\includegraphics[scale=.7,angle=-90]{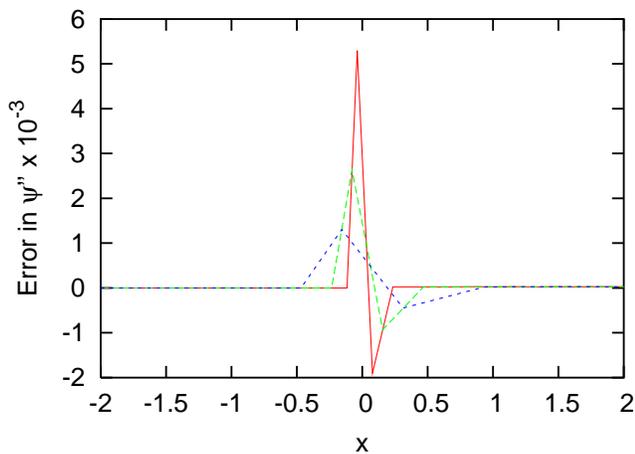}
\caption{Convergence test for $\psi''$. The solid curve is the error in 
$\psi''$ for the highest resolution run. As in Fig.~\ref{psifigure}, 
the interface is at $x=0$ and the errors for the middle (dashed line)
 and low (dotted line) resolution 
cases are divided by factors of 4 and 16, respectively. The grid 
points adjacent to the interface do not coincide because $\psi''$ is only 
first order convergent at these points.}
\label{psippfigure}
\end{figure}
Fig.~\ref{psippfigure} shows the results of our convergence test for $\psi''$. 
The spikes at the interface ($x=0$) appear because the two 
grid points adjacent to the interface are only first order convergent. Elsewhere, the plot shows second 
order convergence.

The behavior demonstrated in Fig.~\ref{psippfigure} also occurs in the BSSN
system when we examine the convergence plot for the Hamiltonian constraint.
In graphing the Hamiltonian constraint $H$, we 
are comparing a combination of grid functions that includes second derivatives 
of the BSSN variables 
to the exact analytical solution for $H$, namely, zero. We therefore expect 
spikes to appear at interfaces in 
the convergence plot for the Hamiltonian constraint, and indeed they do 
(see, for example, Figs.~\ref{fig:gKHP-conv} and \ref{fig:ham+mom_1pluslog}).

We wish to emphasize that the lack of second order convergence for second 
spatial derivatives at grid points adjacent to the interfaces
is not due to any error in our code, or shortcoming of the numerical algorithm.
Since the undifferentiated 
variables are second order convergent everywhere, we can always assure second 
order convergence of their derivatives by using suitable finite difference 
stencils.  For example, in 
computing $D^2\psi^n_j$ from $\psi^n_j$ we can use a second order accurate 
one-sided operator $D^2$ that 
avoids using guard cell values altogether. With such a choice the spikes in 
Fig.~\ref{psippfigure} disappear, 
and $D^2\psi^n_j$ is everywhere second order convergent. 
In our BSSN code, it is most convenient to compute the Hamiltonian constraint 
using the same 
centered difference operator $D^2$ that we use for the evolution equations. As 
a consequence, spikes appear 
at the grid interfaces in the convergence plots (Fig.~\ref{fig:gKHP-conv}
and \ref{fig:ham+mom_1pluslog}).

\section{Analytic Solution for Geodesically Sliced Schwarzschild}
\label{sec:analytic}
In a numerical simulation, geodesic coordinates are obtained by using unit lapse
and vanishing shift.  This implies that the
grid points will follow geodesic trajectories through the physical
spacetime. We present here a physical 
derivation of the Schwarzchild spacetime metric 
in this well-known coordinate system, based on those geodesics; an alternate
derivation is available in Refs. \cite{Misner73,Bruegmann96,Bruegmann97}. 

The Schwarzschild geometry in standard coordinates is given by
\begin{equation}\label{Sch.metric.I}
ds^2=- g_{TT}^{\mbox{ }} dT^2+ g_{RR}^{ }dR^2+R^2d\Omega^2,
\end{equation}
where $g_{TT}^{ } = g_{RR}^{-1}=\left(1-2M/R\right)$.

To express this metric in geodesic coordinates, consider a spatial
Cauchy surface $\Sigma_0$ in a 4-manifold $\M$ and a congruence of
radial geodesics crossing $\Sigma_0$.  Let the affine parameter $\tau$
for each geodesic be zero at $\Sigma_0$. Considering subsequent slices
of constant proper time, we can set a global time $\tau$ which we use 
to define a new foliation $\Sigma_\tau$ of $\M$.  Each geodesic in the
congruence is labeled by the coordinates of its initial ``starting''
point in $\Sigma_0$.  The radial position $\rho$ of the starting point
in $\Sigma_0$ can thus be promoted to a new radial coordinate on $\M$
to pair with the time coordinate $\tau$.

Now we will derive the metric components in this $\tau$-$\rho$
coordinate system. The affine parameter $\tau$ induces the normalized
vector $n^a=(\del/\del\tau)^a$ tangent to the geodesic, implying that the lapse
$g_{\tau\tau}=-1$.  Assuming the geodesics begin at rest so that $n^a$ 
is normal to $\Sigma_0$ implies that $g_{\tau\rho}=0$ initially. 
Furthermore, the geodesic equation $n^a\nabla_a n^b=0$ requires that $g_{\tau\rho,\tau}=0$. 
Thus $g_{\tau\rho}$ (the shift) must remain zero.

A straightforward transformation from Eq.~(\ref{Sch.metric.I}) for the remaining metric coefficient yields 
\begin{equation}
g_{\rho\rho}=
\frac{\left(\del R/\del\rho\right)^2}{\left[\left(\del T/\del\tau\right)g_{TT}\right]^2}.
\end{equation}
The term in the denominator is the energy defined for geodesics on this spacetime, and it is conserved along the geodesics: $n^a\nabla_a(n_b\xi^b)=0$, where $\xi^b=(\del/\del T)^b$ is the timelike Killing field. On $\Sigma_0$ one can
evaluate this energy as $\sqrt{-g^0_{TT}}$, 
where $g^0_{ab}=g_{ab}|_{T=0}$.  This gives
\begin{equation}
g_{\rho\rho}=g^0_{RR}\left(\frac{\del R}{\del\rho}\right)^2=\left(1-\frac{2M}{\rho}\right)^{-1}\left(\frac{\del R}{\del\rho}\right)^2.
\end{equation}

A similar application of conservation of energy in
$n^an_a=-1$ 
yields \cite{Bruegmann96,Bruegmann97}:  
\begin{equation}
\tau-\frac{\rho^{3/2}}{(2M)^{1/2}}\left[ 
\sqrt{\frac{R}{\rho}\left(1-\frac{R}{\rho}\right)}
+ \arccos\sqrt{\frac{R}{\rho}} \right] = 0. 
\label{eq:Geo.r.tau}
\end{equation}
This expression provides an implicit definition for $R=R(\rho,\tau)$, which is easily inverted numerically to high precision.  

To perform numerical evolutions the geodesic coordinates have a
drawback: the physical singularity is already present on the initial
slice $(\tau=0)$ at $\rho=0$. We can avoid this problem by going to
isotropic coordinates $(r,\theta,\phi)$ by means of the transformation
$\rho=r\left(1+{M}/{2r}\right)^2$.  We see that 
\(\rho \rightarrow \infty\) both as \(r \rightarrow 0\) and as
\(r \rightarrow \infty\). For real $r$ the minimum value of $\rho$ is
$\rho=2M$ (the horizon) at $r={M}/{2}$, now the surface closest to
the physical singularity on the initial slice. Substituting 
$\rho = 2M$ into
Eq.~(\ref{eq:Geo.r.tau}) we see that geodesics originating on this
surface reach the physical singularity, $R=0$, at time $\tau=\pi M$,
defining the maximum temporal extent of our coordinate system.

Returning to the metric, the transformation to isotropic coordinates
gives us
\begin{equation}
g_{\rho \rho}=\left(\frac{\del R}{\del \rho}\right)^2  \left(1+\frac{M}{2r}\right)^4.
\label{eq:gr0r0}
\end{equation}
From Eq.~(\ref{eq:gr0r0}) we can express the final metric as:
\begin{equation}
ds^2=-d\tau^2+\left(\frac{\del R}{\del \rho}\right)^2\left(1+\frac{M} {2r}\right)^4
d r^2 + R^2 d\Omega^2.
\label{Ggeod}
\end{equation}

Expressions for the extrinsic curvature, which have not previously
appeared in the literature, can be derived in
a similar manner.
As we know from the ADM formalism \cite{York79,Arnowitt62}, 
the extrinsic curvature can be viewed as the rate of change of
the spatial metric
\begin{equation}
K_{ab}=-\frac{1}{2} \frac{d g_{ab}}{d \tau}
\end{equation}
when the lapse
is unity and the shift is zero.
This gives
\begin{equation}
K_{\rho \rho}=-\frac{1}{2} \left(1+\frac{M} {2r}\right)^4 \frac {\del} {\del\tau} \left(\frac{\del R}{\del
\rho}\right)^2.
\label{Kgeod}
\end{equation}

To evaluate the partial derivatives in Eqs.~(\ref{Ggeod}) and~(\ref{Kgeod}), 
we note that if we have a function $f=f(u,v,w)=0$ defining $u$ as 
an implicit function of $v$ and $w$, we can use the chain rule and the 
implicit function theorem to show that
\begin{eqnarray}
\frac{\partial u}{\partial v} & = & 
-\frac{\partial f/\partial v}{\partial f/\partial u} \\
 \frac{\del^2u}{\del w\del v} &= & -\left(\frac {\del f} {\del u}\right)^{-1} 
 \left[ \frac{\del^2 f} {\del v \del w}  + \frac {\del^2 f}{\del u^2 }
\frac {\del f}{\del v} \frac {\del f}{\del w} \left(\frac {\del f}{\del
u}\right)^{-2}\right]
 \nonumber \\
& & \mbox{} + \left(\frac {\del f}{\del u}\right)^{-2} 
\left(\frac{\del^2 f}{\del v \del u}\frac{\del f}{\del w} + \frac{\del^2
f}{\del u \del w}\frac {\del f}{\del v}\right). \label{eqn:implicitx}
\end{eqnarray}
Taking for $f$ the left hand side of Eq.~({\ref{eq:Geo.r.tau}}), and,
noting that $\del f/\del {\tau} =1$, we conclude
that:
\begin{eqnarray}
K_{\rho \rho} & = & \left( 1 + \frac{M}{2r} \right)^{4} 
\frac{\partial f}{\partial \rho} 
\left( \frac{\partial f}{\partial R} \right)^{-3} \nonumber \\
& & \mbox{} \times \left[ \frac{\partial^{2}f}{\partial \rho \partial R}
- \frac{\partial^{2}f}{\partial R^{2}} \frac{\partial f}{\partial \rho}  
\left( \frac{\partial f}{\partial R} \right)^{-1} \right]
\end{eqnarray}
and
\begin{equation}
K_{\theta\theta}=R \left(\frac{\del f}{\del R}\right)^{-1}  \qquad
K_{\phi\phi}=K_{\theta\theta}\sin^2(\theta). 
\end{equation}
There are no off-diagonal terms. Observe that we have only
partial derivatives of $f$ that can be obtained analytically from
Eq.~(\ref{eq:Geo.r.tau}), and easily evaluated numerically.

\section{Definition of $L_n$ norms and scaling properties}
\label{sec:norms}

For a function $f$ defined on a uniform 
grid $\Delta x = \Delta y = \Delta z \equiv h$,
we take the $L_n$ norm of the function to be
\begin{equation}
L_n[f] = \left(  \sum_{\rm grid} h^3 |f_{jkm}|^n
\right )^{1/n}
\label{eq:Ln-uni}
\end{equation}
where $f_{jkm}$ is the value of the function at grid point $(j,k,m)$;
\emph{cf.}\ \cite{Gustafsson95}.
If the function is defined on a non-uniform grid with 
$\ell$ refinement levels, the $L_n$ norm becomes
\begin{equation}
L_n[f] = \left( 
\sum_{\rm grid\,1} h_1^3 |f_{jkm}|^n + \ldots 
+ \sum_{{{\rm grid}\;\ell}} h_2^3 |f_{jkm}|^n \right )^{1/n},
\label{eq:Ln-FMR}
\end{equation}
where $h_i$ is the cell size on the $i^{\rm th}$ grid.
In our work, the function $f$ denotes an error, either derived
from comparison with an analytic solution (for example, the Hamiltonian
constraint for all our runs, and the basic variables with geodesic
slicing)
 or from comparison with a run at a different resolution as part of
a three-point convergence test (for the basic variables with
1 + log slicing).

It is useful to work out the scaling behavior expected when error
norms from runs with different resolutions are compared.  Recall that,
for the runs presented in this paper, $h_{\rm low} = 2h_{\rm med} = 
4h_{\rm high}$, and the errors in basic variables
such as $\tilde\gamma_{pq}$ and $\tilde A_{pq}$ are expected to scale
as $f \sim h^2$ everywhere.  
 Let $N$ be the characteristic number of zones along one dimension of
a simulation volume, so that $h \sim 1/N$.
We focus on the $L_1$ and $L_2$ norms, which are the ones generally
used to examine errors in numerical relativity.  Then
\begin{equation}
L_1 \sim  \sum_{\rm grid} h^3 f \; \sim \; 
N^3h^3f \; \sim f \;  \sim h^2
\label{eq:L1-scale-basic.vars}
\end{equation}
and
\begin{equation}
L_2 \sim \left ( \sum_{\rm grid}h^3f^2 \right )^{1/2}
 \sim \; f \; \sim h^2,
\label{eq:L2-scale-basic.vars}
\end{equation}
so that both the $L_1$ and $L_2$ norms should exhibit second
order convergence in this case.
Note that these expressions are valid not only for unigrid runs but
also for our FMR simulations, since the refinement structure of the
grid is the same in all these runs.

This situation regarding the $L_1$ and $L_2$ norms of the Hamiltonian
constraint $H$ is somewhat more complicated.  As we have shown in
Sec.~\ref{sec:BHevol} and Appendix~\ref{sec:errors}, $H$ has errors 
that scale as $f \sim h$
on refinement boundaries and as $f \sim h^2$ in the bulk.
For the runs with geodesic slicing, the errors in $H$ in the bulk
near the puncture dominate over those at the refinement boundaries;
see Fig.~\ref{fig:gKHP-conv}.  Since these errors show second
order convergence $f \sim h^2$, we expect that both the $L_1$
and $L_2$ norms will also scale $\sim h^2$, as in 
Eqs.~(\ref{eq:L1-scale-basic.vars}) and~(\ref{eq:L2-scale-basic.vars}).  
However, in the case of 1 + log slicing, the first order convergent
errors on the refinement boundaries play a larger role; see
Fig.~\ref{fig:ham+mom_1pluslog}. To account for this, we write
\begin{equation}
L_n \sim  \left ( \sum_{\rm boundary} h^3 f^n
+ \sum_{\rm bulk} h^3 f^n \right )^{1/n}.
\label{eq:Ln-bound+bulk}
\end{equation}
The number of zones on the boundary $\sim N^2$ while the number
of zones in the bulk $\sim N^3$, for sufficiently large $N$.  Then
\begin{equation}
L_n \sim \left ( N^2h^3h^n + N^3h^3h^{2n}
\right )^{1/n}  \sim (h^{n+1} + h^{2n})^{1/n}.
\label{eq:Ln-bound+bulk2}
\end{equation}
This gives 
\begin{equation}
L_1 \sim h^2
\label{L1-boundaries+bulk}
\end{equation}
and, since $h \ll 1$,
\begin{equation}
L_2 \sim h^{3/2}
\label{L2-boundaries+bulk}
\end{equation}
for the scaling of the norms of $H$ in 1 + log slicing.

\bibliography{references}

\end{document}